\documentclass[runningheads]{llncs}
\makeatletter
\newcommand{\@chapapp}{\relax}%
\makeatother
\usepackage{tikz}
\usetikzlibrary{arrows.meta,decorations.pathreplacing}
\usetikzlibrary{decorations.markings}
\usetikzlibrary{shapes}
\usepackage{algorithm,algpseudocode,float}
\usepackage{xcolor,colortbl}
\usepackage{color,soul}
\usepackage{multirow}
\usepackage{hhline}
\makeatletter
\renewcommand{\fnum@figure}{Figure \thefigure}
\makeatother
\usepackage{amsmath,mathtools}
\usepackage{amssymb}

\usepackage{amsthm}
\usepackage{color,soul}
\newtheorem{thm}{Theorem}
\theoremstyle{definition}
\newtheorem{defn}[thm]{Definition}
\newtheorem{prop}[thm]{Proposition}
\newtheorem{lem}[thm]{Lemma}
\newcommand{\bprop}{\begin{prop}}
\newcommand{\eprop}{\end{prop}}

\newcommand{\bit}{\begin{itemize}}
\newcommand{\eit}{\end{itemize}}
\newcommand{\bcor}{\begin{cor}}
\newcommand{\ecor}{\end{cor}}
\newcommand{\beq}{\begin{equation}}
\newcommand{\eeq}{\end{equation}}
\newcommand{\beqn}{\begin{equation*}}
\newcommand{\eeqn}{\end{equation*}}
\newcommand{\bea}{\begin{eqnarray}}
\newcommand{\eea}{\end{eqnarray}}
\newcommand{\bean}{\begin{eqnarray*}}
\newcommand{\eean}{\end{eqnarray*}}
\newcommand{\ben}{\begin{enumerate}}
\newcommand{\een}{\end{enumerate}}
\newcommand{\bdefn}{\begin{defn}}
\newcommand{\edefn}{\end{defn}}
\newcommand{\bnote}{\begin{note}}
\newcommand{\enote}{\end{note}}
\newcommand{\blem}{\begin{lem}}
\newcommand{\elem}{\end{lem}}
\newcommand{\bthm}{\begin{thm}}
\newcommand{\ethm}{\end{thm}}
\newcommand{\bconj}{\begin{conj}}
\newcommand{\econj}{\end{conj}}
\newcommand{\bconstr}{\begin{constr}}
\newcommand{\econstr}{\end{constr}}
\newcommand{\bpf}{\begin{proof}}
\newcommand{\epf}{\end{proof}}

\newcommand\etc{etc\@ifnextchar.{}{.\@}}

\DeclareMathOperator{\diam}{diam}
\newcommand\ceil[1]{\lceil#1\rceil}
\newcommand\floor[1]{\lfloor#1\rfloor}
\newcommand\card[1]{\lvert#1\rvert}
\newcommand{\etal}{\textit{et al}.}

\newcommand{\orient}[1]{\overrightarrow{#1}}
\def\SNK{\ensuremath{{S_{n, k}}}}
\def\USNK{\ensuremath{\orient{S_{n, k}}}}
\def\SN{\ensuremath{S_n}}
\def\USN{\ensuremath{\orient{S_n}}}

\begin{document}

\title{On Oriented Diameter of $(n, k)$-Star Graphs}
\author{K. S. Ajish Kumar\inst{1}
Birenjith Sasidharan\inst{2} 
K. S. Sudeep\inst{3}}
\authorrunning{K. S. Ajish Kumar, Birenjith Sasidharan, K. S. Sudeep}

\institute{Department of Electronics and Communication Engineering, National Institute of Technology Calicut, India \and
Department of Electronics and Communication Engineering, Government Engineering College Barton Hill, India \and
Department of Computer Science and Engineering, Gandhi Institute of Technology and Management (GITAM) Hyderabad, India, skadavil@gitam.edu}
\maketitle

\begin{abstract}

Assignment of one of two possible directions to every edge of an undirected graph $G=(V,E)$ is called an \emph{orientation} of $G$. The resulting directed graph is denoted by $\orient{G}$. A \emph{strong} orientation is one in which every vertex is reachable from every other vertex via a directed path in $\orient{G}$. The {\em diameter} of $\orient{G}$, i.e., the maximum distance from any vertex to any other vertex, depends on the particular orientation. The minimum diameter among all possible orientations of $G$ is called the \emph{oriented diameter} $\orient{\diam}(G)$ of $G$. Let $n,k$ be two integers such that $1 \leq k < n$. In the realm of interconnection networks of processing elements, an $(n,k)$-star graph \SNK~offers a topology that permits to circumvent the lack of scalability of $n$-star graphs $\SN$. The oriented diameter quantifies an upper limit on the delay in communication over interconnection networks. In this paper, we present a strong orientation scheme for \SNK~that combines approaches suggested by Cheng and Lipman~\cite{cheng2002unidirectional} for \SNK~with the one proposed by Fujita~\cite{fujita2013oriented} for \SN,~reaping benefits from both worlds. Next, we propose a distributed routing algorithm for $\orient{S_{n,k}}$ inspired by an algorithm proposed in \cite{kumar2021oriented} for $\orient{S_n}$. With the aid of both the orientation scheme and the routing algorithm, we show that $\orient{\textsl{diam}}(\SNK)   \leq  \lfloor \frac{n+k}{2} \rfloor + 2k + 6 - \delta(n,k)$ where $\delta(n,k)$ is a non-negative function. The function $\delta(n,k)$ takes on values $2k-n$, $0$, and $\left\lfloor \frac{n-3k}{2} \right\rfloor$ respectively for three disjoint intervals $k>\frac{n}{2}$, $\frac{n}{3} < k \leq \frac{n}{2}$ and $k\leq \frac{n}{3}$. For every value of $n$, $k$, our upper bound performs better than all known bounds in literature. 

\end{abstract}

\section{Introduction}

Let $G=(V,E)$ be an undirected graph with vertex set $V$ and edge set $E$. Assignment of one of two possible directions to every edge of an undirected graph $G=(V,E)$ is called an \emph{orientation} of $G$. Let the resultant directed graph be denoted by $\orient{G}$. There are $2^{\vert E \vert}$ possible orientations for $G$. An orientation is said to \emph{strong orientation} if $\orient{G}$ has at least one directed path between every pair of vertices.
The directed distance  from a node $u$ to $v$ in $\orient{G}$, denoted by $\orient{d}(u,v)$, is the number of edges along the shortest directed path connecting $u$ to $v$. The diameter of $\orient{G}$ is the maximum distance over all pairs of vertices $u,v$ in $\orient{G}$ and is denoted by $\diam(\orient{G})$. In other words, $\diam(\orient{G}) = \max \{\orient{d}(u,v) \vert u,v \in V\}$. The minimum diameter among all possible orientations of $G$ is called \emph{oriented diameter} of $G$. It is denoted by $\orient{\diam}(G)$. Thus $\orient{\diam}(G)= \min \{\diam(\orient{G}) \mid \orient{G} \text{ is a strong orientation of } G\}$. 

Orientation and oriented diameter of graphs are significant in many real-life applications like road transport networks \cite{robbins1939theorem} and interconnection networks in parallel computing. The famous \emph{one-way street problem} is an application of strong orientation in road transport networks. In $1939$, H. E. Robbins \cite{robbins1939theorem} proved that a necessary and sufficient condition for the existence of strong orientation in $G$ is that $G$ remains connected even after the removal of any one of edges. A generalization of the \emph{one-way street} problem addressing strong orientation in mixed multigraph is addressed by F. Boesch and R. Tindell in~\cite{boesch1980robbins}. They established that an undirected edge $(u,v)$ in a mixed multigraph $G$ can be oriented without sacrificing strong orientation if and only if removal of $(u,v)$ does not disconnect $G$. Another notable work in the related area was carried out by Chv\'{a}tal and Thomassen \cite{chvatal1978distances}. They investigated the impact of  orientation of undirected graph $G$ on the distance between two vertices in the oriented graph $\orient{G}$. They showed that  every undirected graph $G$ admits an orientation $\orient{G}$ satisfying the following property: if an edge $(u,v)$ is part of a cycle of length $k$ in $G$ then either $(u,v)$ or $(v,u)$ belong to a cycle of length $(k-2)2^{\floor{\frac{(k-1)}{2}}}+2$ in $\orient{G}$. They also proved that it is NP-hard to decide whether an undirected graph possesses an orientation with diameter at most $2$. Furthermore, they showed that every $2$-edge connected undirected graph of diameter $d$ will possess an orientation with diameter at most $2d^2+2d$. In a recent work \cite{babu2020improvement}, Jasine Babu \etal~ has improved the bound to $1.373d^2+6.971d-1$. 

Strong orientation and oriented diameter of graphs have been studied in detail for many other graph-families such as planar graphs \cite{eggemann2009minimizing}, chordal graphs \cite{fomin2004complexity}, torus \cite{konig1998diameter}, hypercubes \cite{chou1990uni}, $n$-star graphs \cite{day1993unidirectional,fujita2013oriented,kumar2021oriented}, and $(n,k)$-arrangement graphs~\cite{article}. In \cite{cheng2002unidirectional,cheng2006routing}, the problem is investigated with specific focus on $(n,k)$-star graphs,~\SNK. 
The $(n,k)$-star graph is proposed to overcome the scalability issue associated with the $n$-star graphs \SN. 
The $n$-star graph \SN~consists of $n!$ nodes, each of them labelled with a unique permutation of $[n]$. As a result, $S_{n+1}$ contains $n \times n!$ more nodes than that in $S_n$. This huge difference in the number of nodes raises practical concern of resource-wastage in interconnection networks. In this paper, we consider strong orientation in $(n,k)$-star graphs and its oriented diameter.

\subsection{The $(n,k)$-Star Graph \SNK}
\begin{figure}[ht]
	\centering
	\scalebox{.75}{
\begin{tikzpicture}
\tikzset{middlearrow/.style={
        decoration={markings,
        	mark= at position .5 with {\arrow{#1}} ,
        },
        postaction={decorate}
    }
}
\tikzset{endarrow/.style={
        decoration={markings,
        	mark= at position .5 with {\arrow{#1}} ,
        },
        postaction={decorate}
    }
}
\tikzstyle{every node}=[draw=black, ellipse, minimum width=50pt,
    align=center]
    \tikzset{
    ultra thin/.style= {line width=0.1pt},
    ultra thick/.style={line width=1.6pt},
    ultra thicknew/.style={line width=1pt}
    state/.style={circle,inner sep=0pt, minimum size=0pt}
}
\node[ultra thicknew] (a) {7,2,3,4,5}; 
\node[ultra thin,left=130pt, below=55pt] (b) at (a) {5,2,3,4,7};
\node[ultra thin,left=150pt, below=15pt] (c) at (a) {4,2,3,7,5};

\node[ultra thin,left=150pt, above=15pt] (d) at (a) {3,2,7,4,5};
\node[ultra thin,left=130pt, above=55pt] (e) at (a) {2,7,3,4,5};
\node[ultra thick,right=65pt,above=60pt] (f) at (a) {1,2,3,4,5};
\node[ultra thick,right=65pt] (g) at (f) {6,2,3,4,5};
\node[ultra thick,right=195pt] (h) at (a) {8,2,3,4,5};
\node[ultra thick,right=65pt,below=60pt] (i) at (a) {10,2,3,4,5};
\node[ultra thick,right=65pt] (j) at (i) {9,2,3,4,5};

\draw (b) -- node[sloped,font=\small,draw=none,below]{}(a);
\draw (c) -- node[sloped,font=\small,draw=none,below]{}(a);
\draw (a) -- node[sloped,font=\small,draw=none,below]{}(d);
\draw (a) -- node[sloped,font=\small,draw=none,above]{}(e);

\draw (a) -- (g);
\draw (a) -- (h);
\draw (f) -- (a);
\draw (i) -- (f);
\draw (i) -- (a);
\draw (g) -- (i);
\draw (f) -- (h);
\draw (h) -- (i);
\draw (g) -- (f);
\draw (h) -- (g);
\draw (j) -- (a);
\draw (j) -- (f);
\draw (g) -- (j);
\draw (h) -- (j);
\draw (i) -- (j);
\node[ultra thick, below=15pt, label=right:{Clique Node}] (p) at (j){};
\node[ultra thin,below=8pt, label=right:{Star Node}] (q) at (p) {}; 
%

\end{tikzpicture}

}
	\caption{Star neighbours and fundamental clique of the node $7,2,3,4,5$ in $S_{10,5}$}
	\label{nk-star}
\end{figure}
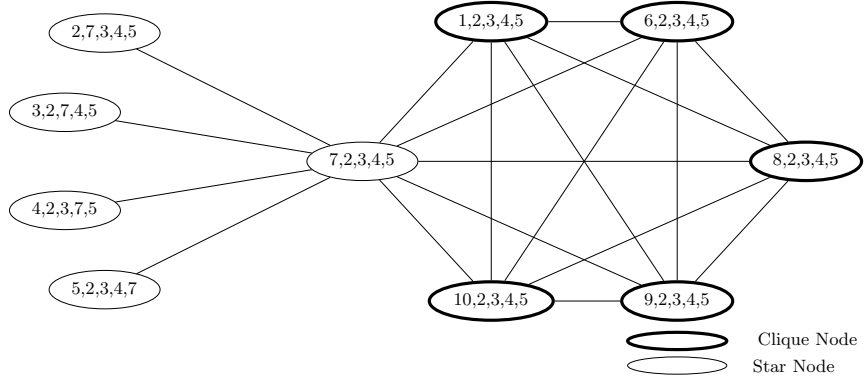

The $(n,k)$-star graph $S_{n,k}$ is determined by two natural numbers $n$ and $k$, where $1 \leq k <  n$. It has $\frac{n!}{(n-k)!}$ nodes and each of them are labelled with a permutation of $k$ elements chosen from the set $\{1,2, \ldots, n\}$. Such a label assigned to a node $u$ is referred to as $k$-permutation label of $u$ denoted by $\sigma_u'$.
A node $\sigma_u'=u_1u_2\ldots u_k$ of \SNK~has two types of neighbours: all nodes of label $xu_2\ldots u_k$, with $x \in \{1,2,\ldots,n\}\setminus \{u_i \mid 1 \leq i \leq k\}$ are called the \emph{clique neighbours}; all nodes of label $u_iu_2\ldots u_{i-1}u_1u_{i+1}\ldots u_k$ (interchanging $u_1$ and $u_i$) with $2 \leq i \leq k$ are called the \emph{star neighbours}. Every node along with its clique neighbours form a complete sub-graph with $(n-k+1)$ nodes. These subgraphs are called \emph{fundamental cliques} and the corresponding edges are referred to as \emph{clique edges}. An $(n,k)$-star graph has $\frac{n!}{(n-k+1)!}$ fundamental cliques. Every node with its star neighbours form a $k$-star sub-graph. These sub-graphs are called \emph{fundamental stars} and the corresponding edges are referred to as \emph{star edges}. There are $\binom{n}{k}$ fundamental stars in $\SNK$. The star neighbours and the fundamental clique of a node $u$ with $\sigma_u'=7,2,3,4,5$ is depicted in Figure~\ref{nk-star}. It is proved in ~\cite{chiang1998topological} that the diameter of $\SNK$ is given by:
\begin{equation} \label{eq:diaSnk}
\textsl{diam}(S_{n,k})=
\begin{cases}
2k-1, & \text{when}\ 1 \leq k \leq \floor{\frac{n}{2}} \\
k+ \floor{\frac{(n-1)}{2}}, & \text{when}\ \floor{\frac{n}{2}} + 1 \leq k < n .
\end{cases} 
\end{equation}

\subsection{Our Contributions}

In the present paper, we propose a strong orientation for \SNK~and a distributed routing algorithm that respects the orientation. With a careful analysis of the routing algorithm, an upper bound on oriented diameter of \SNK~ is derived when $k \geq 3$ and $(n-k) \geq 2$. 

The proposed orientation of \SNK~is a combination of orientations suggested in~\cite{cheng2002unidirectional} and~\cite{fujita2013oriented}. The proposed routing algorithm is arrived at by extending a routing algorithm for $\orient{\SN}$ developed in \cite{kumar2021oriented}. Within every $k$-star subgraph of $S_{n,k}$, our routing algorithm works identical to the one in \cite{kumar2021oriented}. On the other hand, while routing from one $k$-star subgraph to another, we adopt a greedy approach by moving on to one of the best observable directions seen at present node. Putting all these ingredients together, we derive a much tighter upper bound on $\orient{\textsl{diam}}(\SNK)$. In Theorem~\ref{thm:d}, the central theorem of this paper, we prove that 
\bean
\overrightarrow{\textsl{diam}}(S_{n,k}) & \leq & \left\lfloor \frac{n+k}{2} \right\rfloor + 2k + 6 - \delta(n,k) 
\eean
where $\delta(n,k)$ is a non-negative function as given in \eqref{eq:delta}. As $k$ varies in the range $k=1$ to $n-1$, the bound also changes its behaviour. For every $k \leq n/3$, we have $\overrightarrow{\textsl{diam}}(S_{n,k}) \leq 4k+6$ independent the value of $k/n$. For $n/3 < k \leq n/2$, the bound reduces linearly with $k/n$ until it becomes $\overrightarrow{\textsl{diam}}(S_{n,k}) \leq \lfloor 3.5k \rfloor +6$ at $k=\lfloor n/2 \rfloor$. Going further into the regime $n/2 < k \leq n-1$, the bound again reduces linearly with $k/n$ with an alternate slope to reach $\overrightarrow{\textsl{diam}}(S_{n,k}) \leq 2k+7$ at $k=n-1$. In every regime, we achieve significant reduction in comparison with all known upper bounds in literature. 

\section{Preliminaries}

Let $G=(V,E)$ be an undirected graph with vertex-set $V$ and edge-set $E$. The number of edges incident on a vertex $v\in V$ is the degree of $v$. A graph in which all vertices are of equal degree is called a regular graph. A permutation $\sigma$ on $V$ is called an automorphism of $G$ if $\sigma$ maintains the adjacency, i.e., the edge $(u,v)$ is an element of $E$ if and only ($\sigma(u)$,$\sigma(v)$)$\in E$. The graph $G$ is said to be \emph{vertex symmetric} if for any pair of vertices $u,v\in V$, there exists an automorphism that maps $u$ to $v$. The graph $G$ is called \emph{edge symmetric} if for any two edges $(u_1,v_1)$ and $(u_2,v_2)$, there exist an automorphism $\sigma$ such that $\sigma(u_1) = u_2$ and $\sigma(v_1) = v_2$. 


Let $\sigma$ be a permutation of $\{1,2,\ldots,n\}$. We shall write $\sigma = a_1a_2\cdots a_n$ to denote that $\sigma(i) = a_i$. We denote $\sigma^2 = \sigma \circ \sigma$, and in general composition of $\sigma$ to itself $j$ times by $\sigma^j$  for any integer $j > 0$. We also have $\sigma^{-j} = (\sigma^{-1})^j$. The sign of $\sigma$, denoted by $Sign(\sigma)$ is the parity of the number of inversions in $\sigma$, i.e., pairs of values $x$ and $y$ in $\{1,2,\ldots,n\}$ such that $x<y$ and $\sigma(x)>\sigma(y)$. A \emph{cycle} $(a_{j_1},a_{j_2},\ldots,a_{j_m})$ is a permutation $\sigma$ such that $\sigma(a_{j_i})=a_{j_{i+1}}$, $1 \leq i \leq m-1$, $\sigma(a_{j_m}) = a_{j_1}$ and $\sigma(a_i) = a_i$ for every $a_i \notin \{a_{j_{\ell}}, \ell=1,\ldots m \}$. We say a cycle $\psi=(a_{j_1},a_{j_2},\ldots,a_{j_m})$ contains $a$ if $a \in \{a_{j_1},a_{j_2},\ldots,a_{j_m}\}$ and with abuse of notation we may write $a \in \psi$. Two cycles $\psi_1$, $\psi_2$ are disjoint if there does not exist $a \in [n]$ such that $a \in \psi_1$ and $a \in \psi_2$. Every permutation $\sigma$ has a unique decomposition into a composition of disjoint cycles.  Given a permutation $\sigma$, we denote $\psi(a)$ as the cycle containing $a$ in the cyclic decomposition of $\sigma$. Let $a \in [n]$ and we can enumerate the list $a, \sigma(a), \sigma^2(a), \sigma^3(a),\ldots$ in this order and we call this process as {\em traversal in $\psi(a)$ in the forward direction}. In a similar manner, we can enumerate the list $a, \sigma^{-1}(a), \sigma^{-2}(a),\ldots$ in this order and we call this process as {\em traversal in $\psi(a)$ in the backward direction}. Consider a permutation $\sigma=a_1,a_2,\ldots, a_n$ and suppose that we exchange $\sigma(1)$ with $\sigma(k)$ where $k \in \{2,3,\ldots,n\}$ to result in a new permutation $\hat{\sigma}$. When $1$ and $k$ are contained in the same cycle of $\sigma$, then the cycle gets split into two disjoint cycles in $\hat{\sigma}$. When $1$ and $k$ are in two different cycles, then they get merged into a single cycle in $\hat{\sigma}$. We define a relation $\sim_k$ on set of all permutations ${\cal P}_n$ as follows. Let $\sigma_1 = a_1,a_2,\ldots, a_n$ and $\sigma_2 = b_1,b_2,\ldots, b_n$. We say $\sigma_1 \sim_k \sigma_2$ if $\{a_1,a_{k+1},a_{k+2},\ldots,a_{n} \} = \{b_1,b_{k+1},b_{k+2},\ldots,b_{n} \}$. Clearly it is an equivalence relation for any fixed value of $k \leq n$, and it splits ${\cal P}_n$ into a partition of equivalence classes. We call $\sigma_L$ as the leader of a class if it belongs to the class and $\sigma_L(k+1) < \cdots < \sigma_L(n)$. For any permutation $\sigma$, we use $\textsl{lead}(\sigma)$ to denote the leader of the class in which $\sigma$ belongs to.

\section{An Orientation in $(n,k)$-Star Graph}
The $k$-permutation label $\sigma_u'=u_1,u_2,\ldots,u_k$ of a node $u$ can be converted into a permutation on $[n]$ $\sigma_u=u_1,u_2,\ldots u_k,u_{k+1},\ldots u_n$ where the values $u_{k+1},u_{k+2},\ldots,u_n$ are sorted in the ascending order, and form the set $[n] \setminus \{u_1,u_2,\ldots,u_k\}$. The permutation $\sigma_u$ is referred to as the extended permutation label of $u$. By definition, $\sigma_u$ is the leader of an equivalence class defined by the equivalence relation $\sim_k$. We shall use $u$, $\sigma_u'$ or $\sigma_u$ interchangeably to refer a node in $\SNK$. The subsets $\{u_1\}$, $\{u_2,u_3,\ldots, u_{k}\}$ and $\{u_{k+1},u_{k+2},\ldots, u_n \}$ are respectively referred to as \emph{head}, \emph{arm}  and {\em tail-end} of the node $u$. The arm is further split into two subsets $\{u_2,u_3,\ldots, u_{\ceil{\frac{k-1}{2}}+1}\}$ and $\{u_{\ceil{\frac{k-1}{2}}+2},u_{\ceil{\frac{k-1}{2}}+3},\ldots, u_k\}$, respectively referred to as {\em left-half} and {\em right-half} of $u$. The head, arm, left-half, right-half and tail-end of a node $u$ are denoted by $H(u)$, $A(u)$, $L(u)$, $R(u)$ and $T(u)$ respectively. We remark here that the set $L(u)$ and $R(u)$ will be a non-empty set only when $k \geq 3$.

\subsection{Existing Orientations in \SN~and \SNK }

One of the earliest studies on orientation and oriented diameter of \SN~was carried out by Day and Tripathi in \cite{day1993unidirectional}. Much later in $2013$~\cite{fujita2013oriented}, Fujita proposed an orientation scheme for \SN. However, it was proved later by Kumar \etal~\cite{kumar2021oriented} that the scheme considered for \SN~in \cite{day1993unidirectional} is similar in spirit to one suggested by Fujita, though both the schemes may appear different on a peripheral view. In both the schemes, nodes in \SN~are classified into \emph{odd} and \emph{even} nodes based the sign of its label. Recall that the sign of a permutation $\sigma$ is the parity of the minimum number of swaps required to get the identity permutation from $\sigma$. Furthermore, an edge $(u,v)$ is labelled by a number $i$ which is the index of position where the permutation label of $u$ is different from that of $v$ other than the first position. In Day-Tripathi orientation~\cite{day1993unidirectional}, an edge $(u,v)\in \SN$ is oriented from $u$ to $v$, when $u$ is even (odd) signed and parity of edge-label $i$ is even (odd). In Fujita orientation, $(u,v)$ is oriented from  $u$ to $v$ if $u$ is an even (odd) node and the edge-label $i$ is less than (greater than) or equal to $\ceil{\frac{(n-1)}{2}}+1$. Day and Tripathi~\cite{day1993unidirectional} proved that the diameter of \SN~assuming their orientation is $5(n-2)+1$. On the other hand, Fujita proved that the diameter of \SN~ is at most $\ceil{\frac{5n}{2}}+2$. Kumar \etal~\cite{kumar2021oriented} further studied the orientation method by Fujita and proposed a new distributed routing algorithm. With the help of this alternate routing algorithm, they could prove that the diameter of Fujita's orientation is at most $2n+4$. 

In 2002, Cheng and Lipman ~\cite{cheng2002unidirectional} proposed a strong orientation scheme for \SNK~and proved that its oriented diameter is bounded by:
\bean
\orient{\textsl{diam}}(\SNK) & \leq & \left\{ \begin{array}{ll} 
							10k-5 & \text{ when }1 \leq k \leq \floor{\frac{n}{2}}\\
							5k+5\floor{\frac{n-1}{2}} & \text{ when }\floor{\frac{n}{2}}+1 \leq k \leq (n-1)
							\end{array}\right.
\eean 
Later in 2006, Cheng and Kruk ~\cite{cheng2006routing} proposed a distributed routing algorithm for \USNK. They showed that the orientation scheme proposed in ~\cite{cheng2002unidirectional} can lead to an alternate bound for the diameter:
\bean
\orient{\textsl{diam}}(\SNK) & \leq & \left\{ \begin{array}{ll} 
	6(k-3)1+15 &\text{ when }(n-k) \text{ is even}\\
	7(k-3)k+18 &\text{ when }(n-k) \text{ is odd} .
\end{array}\right.
\eean 

\subsection{A Modified Orientation in \SNK \label{sec:orient}}

\begin{figure}[ht]
	\centering
	\scalebox{.75}{\begin{tikzpicture}
\usetikzlibrary{arrows}
\usetikzlibrary{decorations.markings,arrows.meta}
\usetikzlibrary{shapes}

\tikzset{middlearrow/.style={
        decoration={markings,
        	mark= at position .8 with {\arrow{#1}} ,
        },
        postaction={decorate}
    }
}
\tikzset{endarrow/.style={
        decoration={markings,
        	mark= at position .5 with {\arrow{#1}} ,
        },
        postaction={decorate}
    }
}
\tikzset{newarrow/.style={
        decoration={markings,
        	mark= at position .6 with {\arrow{#1}} ,
        },
		ultra thick/.style={line width=1pt},
        postaction={decorate}
    }
}

\tikzstyle{every node}=[draw=black, ellipse, minimum width=50pt,
    align=center]
    \tikzset{
    ultra thin/.style= {line width=0.1pt},
    ultra thick/.style={line width=1.6pt},
    state/.style={circle,inner sep=0pt, minimum size=0pt}
}
\node[ultra thick] (a) {7,2,3,4,5}; 
\node[ultra thin,left=130pt, below=55pt] (b) at (a) {5,2,3,4,7};
\node[ultra thin,left=150pt, below=15pt] (c) at (a) {4,2,3,7,5};
\node[ultra thin,left=150pt, above=15pt] (d) at (a) {3,2,7,4,5};
\node[ultra thin,left=130pt, above=55pt] (e) at (a) {2,7,3,4,5};
\node[ultra thick,right=65pt,above=60pt] (g) at (a) {1,2,3,4,5};
\node[ultra thin,right=80pt] (f) at (g) {6,2,3,4,5};
\node[ultra thin,right=65pt,below=60pt] (h) at (f) {8,2,3,4,5};
\node[ultra thin, right=65pt,below=60pt](j) at (a){10,2,3,4,5};
\node[ultra thick,right=80pt] (i) at (j) {9,2,3,4,5};

\draw[middlearrow={>[scale=3.0]}] (b) -- node[sloped,font=\small,draw=none,below]{5}(a);
\draw[middlearrow={>[scale=3.0]}] (c) -- node[sloped,font=\small,draw=none,below]{4}(a);
\draw[middlearrow={>[scale=3.0]}] (a) -- node[sloped,font=\small,draw=none,below]{3}(d);
\draw[middlearrow={>[scale=3.0]}] (a) -- node[sloped,font=\small,draw=none,above]{2}(e);


\draw[middlearrow={Stealth[scale=2.0]}] (a) -- (g);
\draw[middlearrow={Stealth[scale=2.0]}] (a) -- (h);
\draw[middlearrow={Stealth[scale=2.0]}] (f) -- (a);
\draw[middlearrow={Stealth[scale=2.0]}] (i) -- (a);
\draw[middlearrow={Stealth[scale=2.0]}] (a) -- (j);


\node[ultra thick, right=60pt,below=5pt, label=right:{Even Signed Node}] (p) at (i){};
\node[ultra thin, right=60pt,below=20pt, label=right:{Odd Signed Node}] (q) at (i) {}; 
\node[state, left=40pt,below=35pt](r)at (b){};
\node[state,right=50pt,label=right:{Oriented clique edge}](s) at (r){};
\draw[middlearrow={Stealth[scale=2.0]}] (r) -- (s);

\node[state, left=40pt,below=20pt](t)at (b){};
\node[state,right=50pt,label=right:{Oriented star edge}](u) at (t){};
\draw[middlearrow={>[scale=3.0]}] (t) -- (u);
\end{tikzpicture}}
	\caption{Orientation of edges associated with a node $u$ with $\sigma_{u}=7,2,3,4,5$ in the oriented $S_{10,5}$}
	\label{nk-star-orientation}
\end{figure}
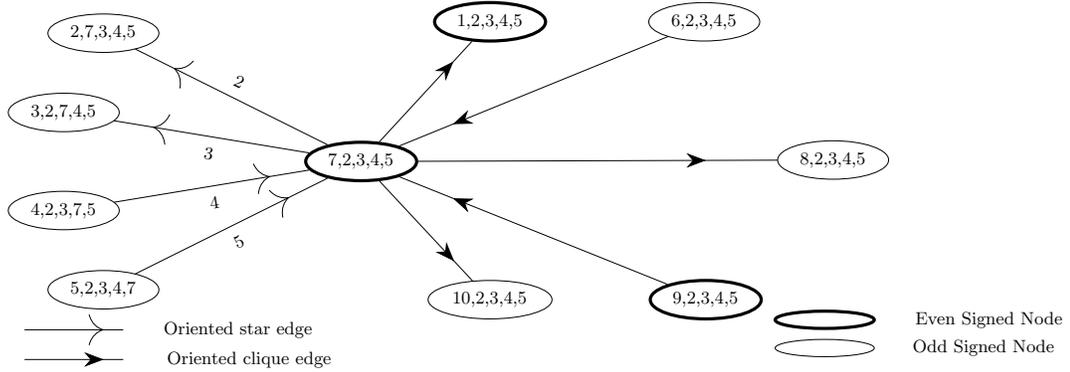

In the proposed orientation, the clique edges are oriented in accordance with the approach in~\cite{cheng2002unidirectional}, whereas the star edges are oriented following Fujita's method for $n$-star graphs~\cite{fujita2013oriented}. 
Every node $u$ in \SNK~is classified as either an \emph{odd} or \emph{even} node based on the sign of its extended permutation label $\sigma_u$. The \emph{star edges} follow the same orientation proposed in~\cite{fujita2013oriented} for an $n$-star graph and the \emph{clique edges} follow the orientation suggested in~\cite{cheng2002unidirectional}. We define the orientation as follows when $k \geq 3$.
\begin{enumerate}
\item Suppose $(u,v)$ is a \emph{star edge}. It is labelled with an integer $i \in \{2, \ldots, k\}$, where the value of $i$ denotes the index of position which differentiate the permutation labels $\sigma_u$, $\sigma_v$. If $\sigma_u$ is even and $i$ belongs to the \emph{left half}, the edge $(u, v)$ is oriented from the node $u$ to $v$. In a similar manner, if $\sigma_u$ is odd and $i$ belongs to the \emph{right half}, the edge $(u, v)$ is oriented from the node $u$ to $v$. Both subsets left half and right half are non-empty as $k \geq 3$. 
\item Suppose the edge $(u,v)$ is a \emph{clique edge}. When the sign of $\sigma_u$ is the same as that of $\sigma_v$, the edge is oriented from $u$ to $v$ if $\sigma_u(1) > \sigma_v(1)$ and $v$ to $u$ if $\sigma_v(1)>\sigma_u(1)$. However, when the sign of $\sigma_u$ is different from that of $\sigma_v$, the edge is oriented from the $u$ to $v$ if $\sigma_u(1) < \sigma_v(1)$ and from the $u$ to $v$ if $\sigma_v(1)<\sigma_u(1)$. 
\end{enumerate}
The oriented star-subgraph and oriented clique-subgraph in which a node $v$ is part of are called \emph{oriented fundamental star} and \emph{oriented fundamental clique} of $v$ respectively. These subgraphs are denoted by $S(v)$ and $Q(v)$. The orientation scheme is illustrated in Figure~\ref{nk-star-orientation} for $\overrightarrow{S_{10,5}}$ taking an example node $v$ with label $\sigma_v'= 7,2,3,4,5$.

\subsection{Structure of Oriented Fundamental Clique in $\protect\orient{S_{n,k}}$\label{sec:ofc}}

First, we will estimate the out-degree of a node in every oriented fundamental clique of $\orient{S_{n,k}}$. 
\bprop \label{prop:directedge} Let $\overrightarrow{S_{n,k}} = (V,E)$, and let $Q(v)$ denote the oriented fundamental clique of $v \in V$. Let $n_E,n_O$ respectively denote the number of even-signed and odd-signed nodes in $Q(v)$. Then $n_E = n_O$ when $(n-k)$ is odd and $|n_E-n_O|=1$ when $(n-k)$ is even. Furthermore, $v$ has at least $\left\lfloor\frac{n-k}{2}\right\rfloor$ outgoing neighbours in $Q(v)$.
\eprop 
\bpf Let the permutation $\sigma$ on $\{1,2,\ldots, n\}$ denote the extended label of node $v$. Let $a_1 < a_2 < \cdots < a_{n-k+1}$ be the ordered listing of $\{\sigma(1),\sigma(k+1),\sigma(k+2),\ldots, \sigma(n) \}$ with $\sigma(1)=a_{\ell}$. Then we can write $\sigma = a_{\ell}, b_2, \cdots , b_k, a_1, \cdots, a_{\ell -1}, a_{\ell +1},  \cdots , a_{n-k+1}$ for some $b_2,b_3,\ldots, b_k \in [n] \setminus \{ a_1,a_{k+1},\ldots, a_{n-k+1} \}$. Let us define a collection of $(n-k+1)$ permutations
\bean
\sigma_i & = & 	a_i, b_2, \cdots  , b_k , a_1 , \cdots , a_{i-1} , a_{i+1} ,  \cdots , a_{n-k+1}
\eean
for $i=1,2,\ldots, n-k+1$. Clearly, $L_Q = \{\sigma_i \mid 1 \leq i \leq n-k+1 \}$ form the set of permutation labels of nodes in $Q(v)$ with $\pi = \sigma_{\ell}$. Given a collection of $(n-k)$ transpositions
\beqn 
t_j = (1 \ k+j)
\eeqn 
for $j=1,2,\ldots , n-k$, it holds that 
\beq
\sigma_{i+1} = t_i \circ t_{i-1} \circ \ \cdots \ \circ t_1 \circ \sigma_1 \label{eq:cycle}
\eeq 
for $i=1,2,\ldots , n-k$. It follows from \eqref{eq:cycle} that for every $i=1,2,\ldots n-k+1$,
\bea
\text{Sign}(\sigma_{i}) & = & \left\{  \begin{array}{cl} \text{Sign}(\sigma_1), &  \text{if $i$ is odd}, \\
	\text{Inverse of } \text{Sign}(\sigma_1), &  \text{if $i$ is even} \end{array} \right. \label{eq:shuffle}
\eea 
Therefore, $n_E = n_O$ when $(n-k)$ is odd and $|n_E-n_O|=1$ when $(n-k)$ is even. 

We will now count the outgoing edges of $v$ labelled by $\sigma_{\ell}$. Let us partition $L_Q \setminus \{\sigma_{\ell}\}$ into $L_{Q,1} = \{ \sigma_{\ell-j} \mid 1 \leq j \leq \ell-1 \}$  and $L_{Q,2} = \{ \sigma_{\ell +j} \mid 1 \leq j \leq n-k+1-\ell \}$. It is clear that $\sigma_j(1) < \sigma_{\ell}(1)$ when $\sigma_j \in L_{Q,1}$ and $\sigma_j(1) > \sigma_{\ell}(1)$ when $\sigma_j \in L_{Q,2}$. We shall count the outgoing edges to $L_{Q,1}$ and $L_{Q,2}$ separately. Outgoing edges from $v$ shall terminate in $L_{Q,1}$ at those permutations with the same sign and in $L_{Q,2}$ at those permutations with the opposite sign as that of $\sigma_{\ell}$. Let us define $\bar{L}_{Q,1} \subset L_{Q,1}$ and $\bar{L}_{Q,2} \subset L_{Q,2}$ as
\bean
\bar{L}_{Q,1} & = & \{ \sigma_{\ell - j} \mid 1 \leq j \leq \ell - 1, j \text{ is even } \} \\
\bar{L}_{Q,2} & = & \{ \sigma_{\ell + j} \mid 1 \leq j \leq n-k+1-\ell, j \text{ is odd } \} . 
\eean
Then $|\bar{L}_{Q,1} \uplus \bar{L}_{Q,2}|$ is the number of outgoing edges of $v$.   
\bean
|\bar{L}_{Q,1} \uplus \bar{L}_{Q,2}| & = & |\bar{L}_{Q,1}| + |\bar{L}_{Q,2}| \\
& = & \left\lfloor \frac{\ell - 1}{2} \right\rfloor + \left\lceil \frac{n-k-(\ell-1)}{2} \right\rceil \\
& \geq &  \left\lfloor \frac{\ell - 1}{2} \right\rfloor + \left( \left\lfloor \frac{n-k}{2} \right\rfloor - \left\lfloor \frac{\ell - 1}{2} \right\rfloor \right) \\
& = & \left\lfloor \frac{n - k}{2} \right\rfloor
\eean
\epf
It is clear by now that the any two nodes $\sigma_1, \sigma_2 \in Q(v)$, we must have $\{\sigma_1(1),\sigma_1(k+1), \ldots, \sigma_1(n)\} = \{\sigma_2(1),\sigma_2(k+1), \ldots, \sigma_2(n)\}$. Furthermore, for any $\sigma \in Q(v)$, $\sigma(k+1) < \sigma(k+2) < \cdots < \sigma(n)$. A node in $Q(v)$ can be alternately represented by a $3$-tuple consisting of a scalar, a set and a vector, i.e., $(j, T(v) \cup H(v),\underline{a}=(\sigma_v(2),\sigma_v(3),\ldots, \sigma_v(k))$ with $j \in H(v)$. Given the $3$-tuple representation, it is straightforward to recover the extended permutation label of the node. When both $T(v) \cup H(v)$ and $\underline{a}$ are specified or clear from the context, a clique node is completely determined by a single number $j$. In that case, we denote the node as $Q(v,j)$. If the nodes are listed as $Q(v,j_1),Q(v,j_2),\ldots, Q(v,j_{n-k+1})$ with $j_1 < j_2 < \cdots < j_{n-k+1}$, then it follows from the proof of Proposition~\ref{prop:directedge} that $Q(v,j_{\ell})$ is an even permutation if and only if $Q(v,j_{\ell+1})$ is odd for every $\ell =1,2,\ldots, n-k$. These notations help to unravel the structure of $Q(v)$ as illustrated with an example below.

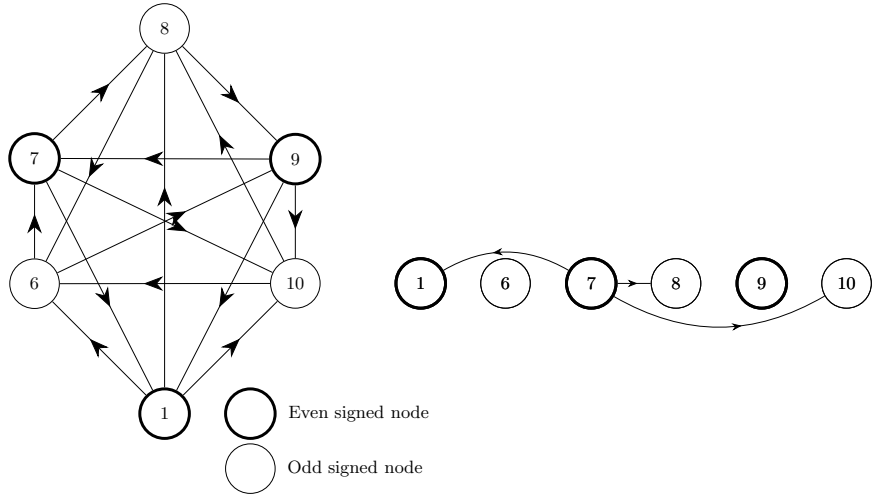
\begin{figure}[ht]
	\centering
	\scalebox{.75}{
%
\begin{tikzpicture}
\usetikzlibrary {positioning}
\usetikzlibrary{arrows}
\usetikzlibrary{decorations.markings,arrows.meta}
\usetikzlibrary{shapes}
\tikzset{middlearrow/.style={
        decoration={markings,
        	mark= at position .6 with {\arrow{#1}} ,
        },
        postaction={decorate}
    }
}
\tikzstyle{every node}=[ultra thick, draw=black, minimum width=25pt,
    align=center]
    \tikzset{
   ultra thick/.style= {circle,
draw,black , line width=1.5pt},
ultra thin/.style= {circle,
draw,black , line width=.5pt},
    ultra rect/.style={rectangle,line width=1.6pt}
}

\node[ultra thick] (a) {1};
\node[ultra thin,above left=80pt] (b) at (a) {6};
\node[ultra thick, above =50pt] (c) at (b) {7};
\node[ultra thin,above right=80pt] (d) at (c) {8};
\node[ultra thick,below right=80pt] (e) at (d) {9};
\node[ultra thin,below =50pt] (f) at (e) {10};
\node[ultra thick,right=50pt] (g) at (f) {1};
\node[ultra thin,right=30pt] (h) at (g) {6};
\node[ultra thick,right=30pt] (i) at (h) {7};
\node[ultra thin,right=30pt] (j) at (i) {8};
\node[ultra thick,right=30pt] (k) at (j) {9};
\node[ultra thin,right=30pt] (l) at (k) {10};

\draw[middlearrow={Stealth[scale=2.0]}] (a)--(b);
\draw[middlearrow={Stealth[scale=2.0]}] (b)--(c);
\draw[middlearrow={Stealth[scale=2.0]}] (c)--(d);
\draw[middlearrow={Stealth[scale=2.0]}] (d)--(e);
\draw[middlearrow={Stealth[scale=2.0]}] (e)--(f);
\draw[middlearrow={Stealth[scale=2.0]}] (a)--(f);
\draw[middlearrow={Stealth[scale=2.0]}] (c)--(a);
\draw[middlearrow={Stealth[scale=2.0]}] (a)--(d);
\draw[middlearrow={Stealth[scale=2.0]}] (e)--(a);
\draw[middlearrow={Stealth[scale=2.0]}] (d)--(b);
\draw[middlearrow={Stealth[scale=2.0]}] (b)--(e);
\draw[middlearrow={Stealth[scale=2.0]}] (f)--(b);
\draw[middlearrow={Stealth[scale=2.0]}] (c)--(f);
\draw[middlearrow={Stealth[scale=2.0]}] (e)--(c);
\draw[middlearrow={Stealth[scale=2.0]}] (f)--(d);

\node[ultra thick,right=50pt] (g) at (f) {1};
\node[ultra thin,right=30pt] (h) at (g) {6};
\node[ultra thick,right=30pt] (i) at (h) {7};
\node[ultra thin,right=30pt] (j) at (i) {8};
\node[ultra thick,right=30pt] (k) at (j) {9};
\node[ultra thin,right=30pt] (l) at (k) {10};
\path[every node/.style={font=\sffamily\small}]
(i) edge[middlearrow={Stealth[scale=1.0]},bend right] node [left] {} (l)
(i) edge[middlearrow={Stealth[scale=1.0]},bend right] node [left] {} (g);
\draw[middlearrow={Stealth[scale=1.0]}] (i)--(j);
\node[ultra thick, right=30pt, label=right:{\,Even signed node}] (p) at (a){};
\node[ultra thin, below=15pt, label=right:{\,Odd signed node}] (q) at (p) {}; 

\end{tikzpicture}
}
	\caption{Illustration of the oriented fundamental clique $Q(v), \sigma_v= 7,2,3,4,5,1,6,8,9,10$ in $S_{10,5}$.}
	\label{fig:Qv}
\end{figure}

Let us consider the same example of $S_{10,5}$ as in Sec.~\ref{sec:orient}. Consider the node $v$ with labels $\sigma_v'= 7,2,3,4,5$ and $\sigma_v= 7,2,3,4,5,1,6,8,9,10$. The nodes in $Q(v)$ are $ Q(v,1),Q(v,6),Q(v,7),Q(v,8),Q(v,9)$ and $Q(v,10)$. Observe that $v=Q(v,7)$. As the context is clear, the nodes may as well be represented by scalars $1,6,7,8,9,10$ with $7$ corresponding to $v$. In Figure~\ref{fig:Qv}, $Q(v)$ is depicted with the help of this notation. The neighbourhood of  $v$ is described arranging all nodes on a line, in ascending order of the scalar values. Since the sign of every permutation alternates in this arrangement, it helps to identify the outgoing neighbours of every node quite easily. For instance, $Q(v,7)$ has outgoing neighbours $Q(v,1), Q(v,8)$ and $Q(v,10)$. What is illustrated in this example generalizes in a straightforward manner. The connectivity in $Q(v)$ is made explicit in the following lemma proved in \cite{cheng2002unidirectional}. The lemma forms the basis for the design and analysis of Algorithm~\ref{CliqueMove} as will be clear later.

\blem \label{lem:3cycle} \cite{cheng2002unidirectional}~Suppose $(n-k)\geq 2$. Let $Q(v)$ be an oriented fundamental clique of node $v$ in $S_{n,k}$ with the orientation proposed in Sec.~\ref{sec:orient}. Let $j_{\text{min}} = \min\{ j \mid Q(v,j) \text{ is a node in } Q(v) \}$ and $j_{\text{max}} = \max\{ j \mid Q(v,j) \text{ is a node in } Q(v) \}$. 
\begin{enumerate} 
\item When $(n-k)$ is even, every directed arc $(Q(v,j_i),Q(v,j_{\ell}))$ (or $(Q(v,j_{\ell}),Q(v,j_{i}))$ as the case may be) belongs to a directed $3$-cycle.
\item When $(n-k)$ is odd, every directed arc $(Q(v,j_{i}),Q(v,j_{\ell}))$ (or $(Q(v,j_{\ell}),Q(v,j_{i}))$ as the case may be) belongs to a directed $3$-cycle except one arc $(Q(v,j_{\text{min}}),Q(v,j_{\text{max}}))$. It belongs to a directed $4$-cycle.
\end{enumerate}
\elem 
\bpf The proof is by constructing suitable cycles. Let $Q(v,j_1),Q(v,j_2), \ldots,$ $Q(v,j_{n-k+1}), j_1 < j_2 < \ldots < j_{n-k+1}$ be the nodes in $Q(v)$. Consider a directed arc $(Q(v,j_i),Q(v,j_{\ell}))$ in $Q(v)$. 

First, let us consider the case when $(n-k)$ is even. Suppose $i<\ell$. Then the extended permutation labels of $Q(v,j_i)$ and $Q(v,j_{\ell})$ are of opposite sign. When $i \neq 1$, we have a directed $3$-cycle: $(Q(v,j_i)\rightarrow Q(v,j_{\ell})\rightarrow Q(v,j_{i-m})\rightarrow Q(v,j_i))$ for some positive odd number $m$ such that $i-m\geq 1$. When $i = 1$, we have a directed $3$-cycle: $(Q(v,j_i)\rightarrow Q(v,j_{\ell})\rightarrow Q(v,j_{\ell +1})\rightarrow Q(v,j_i))$. Suppose $i>\ell$. Then the extended permutation labels of $Q(v,j_i)$ and $Q(v,j_{\ell})$ are of the same sign. The directed arc $(Q(v,j_i),Q(v,j_l))$ is part of a directed $3$-cycle $(Q(v,j_i)\rightarrow Q(v,j_{\ell})\rightarrow Q(v,j_{\ell+1})\rightarrow Q(v,j_i))$.

Next, consider the case when $(n-k)$ is odd. Every argument for even $(n-k)$ remains true here as well. But in addition, we need to consider the arc $(Q(v,j_1),Q(v,j_{n-k+1}))$, which can not exist when $n-k$ is even.  This arc is part of a directed $4$-cycle $Q(v,j_1)\rightarrow Q(v,j_{n-k+1})\rightarrow Q(v,j_{n-k+1-m})\rightarrow Q(v,j_{n-k+2-m})\rightarrow Q(v,j_1)$ for some positive even number $m$ such that $n-k+1-m \geq 2$. 
\epf

%
%
A directed shortest path from $Q(v,x)$ to $Q(v,y)$ in $Q(v)$ is denoted by $P_v(x,y)$. If a node $Q(v,z)$ is visited in the path, then we write $Q(v,z) \in P_v(x,y)$. The $\ell$-th node visited after $x$ on $P_v(x,y)$ starting from $x$ in order is denoted as $P_v(x,y,\ell)$. For example, $P_v(x,y,0)=Q(v,x)$. For the oriented clique $Q(v)$ in the example discussed, three directed paths $P_v(7,10)=Q(v,7),Q(v,10)$, $P_v(7,9)=Q(v,7),Q(v,8),Q(v,10)$ and $P_v(1,10)=Q(v,1),Q(v,6),Q(v,9),Q(v,10)$ can be identified in Figure~\ref{fig:Qv}. While the first and second paths belong to $3$-cycles, the third one belongs to a $4$-cycle.
  
\section{A Routing Algorithm and the Oriented Diameter}

In \cite{kumar2021oriented}, authors mention two ways of looking at a routing algorithm for \USN~ -- the ``network view'' and the ``sorting view''. These two perspectives apply to the case of \USNK~as well. Let $\sigma_s$ and $\sigma_t$ denote permutation labels of source node $s$ and destination node $t$. In the network view, a packet that includes the {\em address} $\sigma_t$ in its header originates at $s$, and every node that receives the packet routes it to one of its neighbour. Depending on $\sigma_t$, it can be routed via a star edge or a clique edge. In the sorting view, transmission of the packet from a node $u$ to its neighbour $v$ can be viewed as a composition of permutation $\sigma_u$ with $\pi_{u,v}$ to result in $\sigma_v = \pi_{u,v} \circ \sigma_u$. If the routing happens via a star-edge, then $\pi_{u,v}$ is a transposition. If the routing happens via a clique-edge, then $\pi_{u,v}$ is composition of multiple transpositions as described in \eqref{eq:cycle}. We shall invoke these two perspectives interchangeably in both the description of routing algorithm and subsequent analyses. We begin with definitions of certain data structures that are relevant to the routing algorithm. 
 
\begin{defn}\label{DataStructures} Let $s$, $t$ be source and destination. Suppose $c$ be a node visited in between. We define the following data structures\footnote{Two data-structures $IA(t)$ and $E(t)$ are redundant as they are precisely $A(t)$ and $T(t)$ respectively. But a slight degree of redundancy aids in giving notation for subsequently defined sets following mnemonic strategy.}:
	 \begin{enumerate}
		\item \emph{Internal values: }  $I(t) = \{\sigma_t(1), \sigma_t(2), \ldots, \sigma_t(k)\}$
		\item \emph{Internal values belonging to the arm: }  $IA(t) = I(t) \cap A(t) = A(t)$
		\item \emph{Displaced internal values: } $DI(c,t) =  I(t) \cap T(c) $
		\item \emph{External values: }  $E(t) = [n] \setminus I(t) = T(t)$
		\item \emph{Displaced external values: } $DE(c,t) = E(t) \cap (H(c) \cup A(c))$
		\item \emph{Displaced external values in the arm: } $DEA(c,t) = E(t) \cap A(c)$
		\item \emph{Displaced external values in the left-half: } $DEL(c,t) = E(t) \cap L(c)$
		\item \emph{Displaced external values in the right-half: } $DER(c,t) = E(t) \cap R(c)$
		\item \emph{Settled values: } $S(c,t) = \{\sigma_c(i) \mid \sigma_c(i) \in I(t), \sigma_c(i) = \sigma_t(i), 1 \leq i \leq k \}$.
		\item \emph{Settled values in the left-half: } $SL(c,t) = S(c,t) \cap L(t)$
		\item \emph{Settled values in the right-half: } $SR(c,t) = S(c,t) \cap R(t)$
		\item \emph{Unsettled values: } $U(c,t) = \{\sigma_c(i) : \sigma_c(i) \in I(t), \sigma_c(i) \neq \sigma_t(i), 1 \leq i \leq k \}$
		\item \emph{Unsettled values in the left-half belonging to the left-half of $t$: }	$ULL(c,t) = U(c,t) \cap L(c) \cap L(t)$ 
		\item \emph{Unsettled values in the right-half belonging to the right-half of $t$: } $URR(c,t) = U(c,t) \cap R(c) \cap R(t)$ 
		\item \emph{Unsettled values in the left-half belonging to the right-half of $t$: } $ULR(c,t) = U(c,t) \cap L(c) \cap R(t)$ 
		\item \emph{Unsettled values in the right-half belonging to left-half of $t$: } $URL(c,t) = U(c,t) \cap R(c) \cap L(t)$
	\end{enumerate}
\end{defn}


In the proposed routing algorithm, given that we have reached an intermediate node $c$, the choice of next node $r$ is solely determined by $c$, and does not depend upon the path taken from $s$ to reach $c$. If $\sigma_c(1)$ is an external value, then choose a neighbour $r$ so that $(c,r)$ is a clique-edge. We call this a clique-move. On the other hand, if $\sigma_c(1)$ is an internal value, then choose a neighbour $r$ so that $(c,r)$ is a star-edge, except possibly when $\sigma_c(1)=\sigma_t(1)$. And we call this a star-move. Whenever $\sigma_c(1)=\sigma_t(1)$, we may do a star-move or clique-move depending on when $DEA(c,t)$ is empty or not, and they are given special names: {\em seeding star-move} and {\em seeding clique-move}. The behaviour of star-move and seeding star-move exactly follows the approach in \cite{kumar2021oriented}. From a sorting-viewpoint, all the moves can be viewed as performing a transposition operation $(1 \ i) \circ \sigma_c$. If it is clique-move, then $i \in \{k+1,k+2,\ldots,n \}$. However the transposition is followed by another set of transpositions so as to re-order the tail-end values. If it is a star-move, then $i \in \{2,3,\ldots,n \}$. Whether $i$ is picked from the left-half or right-half of $c$ depends on the sign of permutation $\sigma_c$. Thus we can summarize the moves as follows:
\begin{itemize}
	\item Seeding clique-move: $\sigma_c(1)$ is same as $\sigma_t(1)$, and $i \in \{k + 1, \ldots , n\}$.
	\item Seeding star-move: $\sigma_c(1)$ is same as $\sigma_t(1)$, and $i \in \{2, \ldots , k\}$.
	\item Clique-move : $\sigma_c(1)$ is an external value, and $i \in \{k + 1, \ldots , n\}$.
	\item Star-move: $\sigma_c(1)$ is an internal value other than $\sigma_t(1)$, and $i \in \{2, \ldots , k\}$.
\end{itemize}

The routing algorithm is presented in Algorithm~\ref{RouteEven}. It invokes four sub-routines associated with the four moves: (i) Algorithm~\ref{SeedClique} for seeding clique-move, (ii) Algorithm~\ref{SeedStar} seeding star-move, (iii) Algorithm~\ref{CliqueMove} for clique-move and (iv) Algorithm~\ref{StarMove} for star-move. When a star-move is executed, it can be either a \emph{settling move} or a \emph{crossing move}~\cite{kumar2021oriented}. If $\sigma_c(1) \in L(t)$ and $\sigma_c$ is even (or $\sigma_c(1) \in R(t)$), and $\sigma_c$ is odd), then the star-move is a \emph{settling move}. On the other hand if $\sigma_c(1) \in  L(t)$ and $\sigma_c$ is odd (or $\sigma_c(1) \in R(t)$ and $\sigma_c$ is even), then the move is a \emph{crossing move}. The sub-routines Algorithm~\ref{SeedStar} and Algorithm~\ref{StarMove} depend on the sign of $\sigma_c$, and we present them assuming $\sigma_c$ is even. If $\textsl{Sign}(\sigma_c)$ is even and a star-move is initiated at $c$, then $\sigma_c(1)$ replaces an element in the left-half $L(c)$. On the other hand, if $\textsl{Sign}(\sigma_c)$ is odd, then $\sigma_c(1)$ replaces an element in the right half $R(c)$. However, every rule that determines which element in either $L(c)$ or $R(c)$ is picked remain the same. Therefore, when $\sigma_c$ is odd, the two subroutines remain the same, except for replacing $L(c),L(t)$, $R(t)$, $ULL(c,t)$, $ULR(c,t)$, $URR(c,t)$, $DEL(c,t)$, and $SL(c,t)$ with $R(c)$, $R(t)$, $L(t)$, $URR(c,t)$, $URL(c,t)$, $ULL(c,t)$, $DER(c,t)$, and $SR(c,t)$ in that order.


\begin{algorithm*}[htp]
\begin{algorithmic}[1]
\caption{Processing done by a node $c$ 
	upon receiving a packet $P$ destined for a node $t$.}\label{RouteEven}
  \Procedure{(n,k)-Star-Route(Packet $P$)}{}
    \State Let $c$ be the current node and $t$ be the destination  
  	\State {\em If} $\sigma_c == \sigma_t$ {\em then}
  	\State \hspace{.2cm} Accept $P$, terminate routing and exit.
    \State Let $E=E(t)$, $DEA = DEA(c,t)$, $I=I(t)$.
    \State {\em Case 1:} $\sigma_c(1)==\sigma_t(1)$ {\em AND} $\card{DEA} >0$
    \State \hspace{.2cm} $r = $ \textsc{Seed-Clique}($c$,$t$).
    \State {\em Case 2:} $\sigma_c(1)==\sigma_t(1)$
    \State \hspace{.2cm} $r = $ \textsc{Seed-Star}($c$,$t$).
    \State {\em Case 3:} $\sigma_c(1) \in E$ 
    \State \hspace{.2cm} $r = $ \textsc{Clique-Move}($c$,$t$). 
    \State {\em Case 4:} $\sigma_c(1) \in I$
    \State \hspace{.2cm} $r = $ \textsc{Star-Move}($c$,$t$). 
 	\State Forward $P$ to the node $r$.
 \EndProcedure
\end{algorithmic}
\end{algorithm*} 

\begin{algorithm}[htp]
\begin{algorithmic}[1]
\caption{Processing done by a node $c$ when $\sigma_c(1)==\sigma_t(1)$ and $\vert DEA(c,t) \vert >0 $.}\label{SeedClique}
  \Procedure{Seed-Clique(Node $c$, Node $t$)}{}
  	\State Let $N_Q(c)$ be the subset of nodes in $Q(c)$ to which $c$ has an outgoing edge.
  	\State Let $X = \{\sigma_x(1): x \in N_Q(c)\}$.
  	\State Let $E=E(t)$, $I=I(t)$.
  	\State {\em Case 1:} $\exists y \in I \cap X$.
  	\State \hspace{.2cm} $i=\sigma_c^{-1}(y)$.
  	\State {\em Case 2:} $\exists y \in E \cap X, \exists z \in I$ such that there is a shortest path from $c$ to $Q(c,z)$ via $Q(c,y)$.
  	\State \hspace{.2cm} $i=\sigma_c^{-1}(y)$.
  	\State $\sigma = (1\, i) \circ \sigma_c$.  	
  	\State $\sigma_{r} = \textsl{lead}(\sigma)$ 
  	\State return $r$.
 \EndProcedure
\end{algorithmic}
\end{algorithm}

\begin{algorithm}[htp]
\begin{algorithmic}[1]
\caption{Processing done by an even node $c$ when $\sigma_c(1)==\sigma_t(1)$ and $\vert DEA(c,t) \vert ==0 $.}\label{SeedStar}
  \Procedure{Seed-Star(Node $c$, Node $t$)}{}
  	\State Let $ULL=ULL(c,t)$, $URR=URR(c,t)$, $SL = SL(c,t)$, $ULR=ULR(c,t)$.
  	\State {\em Case 1:} $\card{ULL}+\card{URR} = 0$
	\State \hspace{.5cm} $i=\sigma_c^{-1}(y)$ such that $y \in ULR$.
	\State {\em Case 2:} $\card{ULL}+\card{URR} > 0$
  	\State \hspace{.5cm} $i=c^{-1}(y)$ such that $y$ is a member of the first non-empty set taken in the order $ULL$, $SL$, $ULR$.
	\State $\sigma_r = (1\, i) \circ \sigma_c$
  	\State return $r$.
 \EndProcedure
\end{algorithmic}
\end{algorithm}

\begin{algorithm}[htp]
	\begin{algorithmic}[1]
		\caption{Processing done by a node $c$ when $\sigma_c(1)\in E(t)$.}\label{CliqueMove}
		\Procedure{Clique-Move(Node $c$, Node $t$)}{}
		\State Let $N_Q(c)$ be the set of nodes in $Q(c)$ to which $c$ has an outgoing edge.
		\State Let $X = \{\sigma_x(1): x\in N_Q(c)\}$, $T=T(c)$.
		\State Let $E=E(t)$, $IA=IA(t)$.
		\State Let $IAX = IA \cap X$, $IAT = IA \cap T$
		\State {\em Case 1:} $\card{IAX} > 0$
		\State \hspace{.5cm} $i=\sigma_c^{-1}(y)$ such that $y \in IAX$.
		\State {\em Case 2:} $\card{IAT} > 0$ 
		\State	\hspace{.5cm} {\em Case 2.1:} $\exists z\in IAT$ such that $P_c(\sigma_c(1),z)$ is a shortest path and $\sigma_t(1) \notin P_c(\sigma_c(1),z)$
		\State \hspace{1cm} $y = P_c(\sigma_c(1),z,1)$
		\State \hspace{1cm} $i = \sigma_c^{-1}(y)$
		\State \hspace{.5cm} {\em Case 2.2:} $\exists z\in IAT$ such that $P_c(\sigma_c(1),z)$ is a shortest path and $\sigma_t(1) \in P_c(\sigma_c(1),z)$
		\State \hspace{1cm} $y = P_c(\sigma_c(1),z,1)$
		\State \hspace{1cm} $i = \sigma_c^{-1}(y)$
		\State {\em Case 3:} $\card{IAT} == 0$
		\State \hspace{.5cm} Let $P_c(\sigma_c(1),\sigma_t(1))$ be a shortest path in $Q(c)$
		\State \hspace{0.5cm} $y = P_c(\sigma_c(1),\sigma_t(1),1)$
		\State \hspace{0.5cm} $i = \sigma_c^{-1}(y)$
		\State $\sigma = (1\, i) \circ \sigma_c$.
		\State $\sigma_r = \textsl{lead}(\sigma)$
		\State return $r$.
		\EndProcedure
	\end{algorithmic}
\end{algorithm}

\begin{algorithm}[htp]
\begin{algorithmic}[1]
\caption{Processing done by an even node $c$ when $\sigma_c(1) \in I(t)$ and $\sigma_c(1) \neq \sigma_t(1)$.}\label{StarMove}
  \Procedure{Star-Move(Node $c$, Node $t$)}{}
  	\State Let $E=E(t)$, $DEL=DEL(c,t) \cap L(c)$, $SL=SL(c,t)$, $ULR = ULR(c,t)$, and  $ULL=ULL(c,t)$.
  	\State {\em Case 1 (Settling Move):} $\sigma_c(1) \in L(t)$.
 	\State \hspace{.5cm} $i=\sigma_t^{-1}(\sigma_c(1))$.
	\State {\em Case 2 (Crossing Move):} $\sigma_c(1) \in R(t)$.
		\State \hspace{.5cm}Let $\psi=\psi_{\sigma_c(1)}$ be the cycle that contains $\sigma_c(1)$ in cyclic decomposition of $\sigma_c$.
		\State \hspace{.5cm}{\em Case 2.1:} $\exists y \in DEL \cup ULL$ such that $\sigma_c(i)\not\in \psi$
 		\State \hspace{1cm} $i=\sigma_c^{-1}(y)$
 		\State \hspace{.5cm}{\em Case 2.2:} $DEL \cup ULL \neq \phi$ and $y \in \psi$ for every $y \in DEL \cup ULL$
 		\State \hspace{1cm} Let $y \in DEL \cup ULL$ is encountered first while traversing $\psi$ backward from $\sigma_c(1)$.
 		\State \hspace{1cm} $i=\sigma_c^{-1}(y)$ 	
 		\State \hspace{.5cm}{\em Case 2.3:} $\exists y \in SL$.
 		\State \hspace{1cm} $i=\sigma_c^{-1}(y)$	
 		\State \hspace{.5cm}{\em Case 2.4:} $\exists y \in ULR$ such that $y \not \in \psi$ and $y$ belongs to an alternating cycle
 		\State \hspace{1cm} $i=\sigma_c^{-1}(y)$ 	
 		\State \hspace{.5cm}{\em Case 2.5:} $\exists y \in ULR$ such that $y \not \in \psi$
 		\State \hspace{1cm} $i=\sigma_c^{-1}(y)$
 		\State \hspace{.5cm}{\em Case 2.6:} $\exists y \in ULR$
  		\State \hspace{1cm} $i=\sigma_c^{-1}(y)$ 
 	\State $\sigma_r = (1\, i) \circ \sigma_c$.
	\State return $r$. 	
 \EndProcedure
\end{algorithmic}
\end{algorithm}

\subsection{Correctness of the Algorithm}

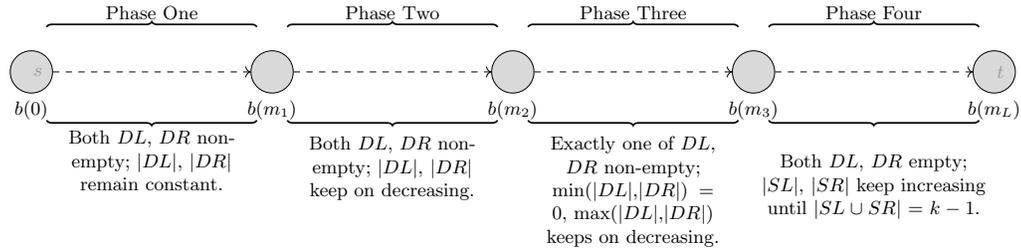
\begin{figure}[ht]
	\centering
	\scalebox{.8}{
%
%
%

\begin{tikzpicture}
[
	mycircle/.style={
         circle,
         draw=black,
         fill=gray,
         fill opacity = 0.3,
         text opacity=1,
         inner sep=0pt,
         minimum size=20pt,
         font=\small},
	state/.style={circle,inner sep=0pt, minimum size=2pt},
    node distance=1.2cm
]
        
  \node[mycircle,label=below:$b(0)$] (p0) {$s$};
  \node[mycircle, right of= p0,node distance = 4cm,label=below:$b(m_1)$](p1) {};
  \node[mycircle, right of= p1,node distance = 4cm,label=below:$b(m_2)$](p2) {};
  \node[mycircle, right of= p2,node distance = 4cm,label=below:$b(m_3)$](p3) {};
  \node[mycircle, right of= p3,node distance = 4cm,label=below:$b(m_L)$](p4) {$t$};
\draw[dashed,->](p0) -- (p1);
\draw[dashed,->](p1) -- (p2);
\draw[dashed,->](p2) -- (p3);
\draw[dashed,->](p3) -- (p4);
\draw [
   thick,
    decoration={
       brace,
       mirror,
       raise=0.55cm
   },
   decorate
] (p0.south east) -- (p1.south west) 
node [pos=0.5,anchor=north,yshift=-0.6cm,align=center,text width=3.8cm,midway,font=\small]{Both $DL$, $DR$ non-empty; $\vert DL\vert$, $\vert DR\vert$ remain constant.};
\draw [
   thick,
    decoration={
       brace,
       mirror,
       raise=0.55cm
   },
   decorate
] (p1.south east) -- (p2.south west) 
node [pos=0.5,anchor=north,yshift=-0.7cm,align=center,text width=3.8cm,midway,font=\small]{Both $DL$, $DR$ non-empty; $\vert DL\vert$, $\vert DR\vert$ keep on decreasing.}; 
\draw [
   thick,
    decoration={
       brace,
       mirror,
       raise=0.55cm
   },
   decorate
] (p2.south east) -- (p3.south west) 
node [pos=0.5,anchor=north,yshift=-0.7cm,align=center,text width=3.8cm,midway,font=\small]{Exactly one of $DL$, $DR$ non-empty;\\ $\min(\vert DL\vert$,$\vert DR\vert)=0$, $\max(\vert DL\vert$,$\vert DR\vert)$ keeps on decreasing.}; 
\draw [
   thick,
    decoration={
       brace,
       mirror,
       raise=0.55cm
   },
   decorate
] (p3.south east) -- (p4.south west) 
node [pos=0.8,anchor=north,yshift=-1cm,align=center,text width=3.8cm,midway,font=\small]{Both $DL$, $DR$ empty;\\ $\vert SL\vert$, $\vert SR\vert$ keep increasing until $\vert SL \cup SR\vert=k-1$.};

\draw [
    thick,
    decoration={
        brace,
        raise=1cm
    },
    decorate
] (p0.south east) -- (p1.south west) 
node [pos=0.5,anchor=south,yshift=1cm] {Phase One};
\draw [
    thick,
    decoration={
        brace,
        raise=1cm
    },
    decorate
] (p1.south east) -- (p2.south west) 
node [pos=0.5,anchor=south,yshift=1cm] {Phase Two};
\draw [
    thick,
    decoration={
        brace,
        raise=1cm
    },
    decorate
] (p2.south east) -- (p3.south west) 
node [pos=0.5,anchor=south,yshift=1cm] {Phase Three};
\draw [
    thick,
    decoration={
        brace,
        raise=1cm
    },
    decorate
] (p3.south east) -- (p4.south west) 
node [pos=0.5,anchor=south,yshift=1cm] {Phase Four}; 
  
\end{tikzpicture}
}
	\caption{History of moves during the routing from $s$ to $t$.}
	\label{fig:Hm}
\end{figure}

\bdefn Suppose a packet is routed from $s$ to $t$ using the proposed routing algorithm. \ben 
\item Execution of either of \textsc{star-move}, \textsc{clique-move}, \textsc{seed-star} or \textsc{seed-clique} is referred to as a move. Execution of each of them in order is termed as star-move, clique-move, seeding star-move and seeding clique-move. The node arrived at after the execution of the $m$-th move is denoted by $b(m)$. Let the last move for routing be $m_L$ and therefore $b(m_L)=t$. By definition, $b(0)=s$. 
\item If $\sigma_{b(m-1)}(1) = x$ and $\sigma_{b(m)}^{-1}(x) = j$, $\sigma_{b(m-1)}(j) = y$ then we say $x$ replaces $y$ in the $m$-th move.
\item Suppose for $j \in [k]$, $\sigma_{b(m-1)}(1) = \sigma_t(j)$ and $\sigma_{b(m)}(j) = \sigma_t(j)$, then we say $\sigma_t(j)$ is settled at the $m$-th move. The $m$-th move is called as settling move for $\sigma_t(j)$.
\item Suppose for $j \in [k]$, $\sigma_t(j) \in L(t)$ (alternately $R(t)$), $\sigma_{b(m-1)}(1) = \sigma_t(j)$ and $\sigma_t(j) \in R(b(m))$ (alternately $L(b(m))$), then we say $\sigma_t(j)$ is crossed at the $m$-th move. The $m$-th move is called as crossing move for $\sigma_t(j)$.
\een
\edefn 
While we have already made a qualitative description on various types of moves (both from network-view and sorting-view) at the beginning of this section, the above definition makes it precise and suitable for carrying out analyses. Let 
\bean
{\cal H} & = & \{ (\sigma_c,\sigma_r) \mid \text{node $r$ is visited next to $c$} \}
\eean
denote the history of all moves that have taken place. We can also define three natural subsets of ${\cal H}$ as 
\bean
{\cal H}_{\text{star}} & = &  \{ (\sigma_c, \ \sigma_r) \mid r \text{ is visited next to $c$ and } r=\textsc{star-move}(c,t) \} \\
{\cal H}_{\text{clique}} & =  & \{ (\sigma_c, \ \sigma_r) \mid r \text{ is visited next to $c$ and } r=\textsc{clique-move}(c,t) \} \\
{\cal H}_{\text{seed}} & = & \{ (\sigma_c, \ \sigma_r) \mid r \text{ is visited next to $c$ and } r=\textsc{seed-clique}(c,t) \text{ or } r=\textsc{seed-star}(c,t) \}
\eean 
The entire history of moves can be partitioned into four phases \textsl{Phase 1--4}(Figure~\ref{fig:Hm}) based on the sizes of $DL(c) = ULL(c,t) \cup DEL(c,t)$ and $DR(c) = URR(c,t) \cup DER(c,t)$, where $c$ is the current node and $t$ is a fixed destination. We choose to keep $DL(c)$ and $DR(c)$ as a function of $c$ alone, as we consider a scenario of routing to a fixed destination $t$. 

In \textsl{Phase 1}, both the sets $DL(c)$ and $DR(c)$ are non-empty and their sizes remain constant as a function of $c$. We call this phase as {\em transient phase}. The transient phase will begin from the first move and end by move $m_1$ where $m_1$ is given by
\bea
m_1 = \max \{ m \mid |DL(b(m)) \cup DR(b(m)) | \ = \ |DL(b(m')) \cup DR(b(m')) | \nonumber\\\text{ for every } m' \leq m \} \label{eq:m1}.
\eea
The number of moves in the transient phase depends on $\sigma_s(1)$. Three possible moves can occur in transient phase: (a) if settling of $\sigma_s(1)$ happens replacing $y \in ULR(s,t)$, then a few more settling star-moves can follow (b) a seeding clique-move can happen if $\sigma_s(1)=\sigma_t(1)$, (c) a clique-move can happen if $\sigma_s(1)\in E(t)$. Since replacement of an element in $DL$ has the highest priority in both crossing star-move and seeding star-move sub-routines, it is bound to happen in finite number of moves, i.e., $m_1 < \infty$ and hence the transient phase will terminate. The $(m_1+1)$-st move witnesses the first star-move leading to reduction in either $|DL|$ or $|DR|$, and we enter \textsl{Phase 2}. In \textsl{Phase 2}, both $|DL|$ and $|DR|$ will invariably keep on reducing until one of them becomes zero. If 
\bea
m_2 & = & \min \{ m \mid m \geq m_1, \text{ either }|DL(b(m))|=0 \text{ or } |DR(b(m))| = 0 \} \label{eq:m2},
\eea
then \textsl{Phase 2} lasts till $m_2$-th move starting from $(m_1+1)$. If $y \in DEL(c,t)$ gets replaced as part of a move from $c$ in \textsl{Phase 2}, then a clique-move follows. On the other hand, if $y \in ULL(c,t)$ is replaced, then a crossing star-move follows. In the former case, a transient behaviour (akin to what happens in transient phase) is bound to happen before the next move to replace an element in $DL$. A symmetric scenario holds true for $DR$. For this reason, \textsl{Phase 2} shall be called {\em symmetric crossing phase}. In \textsl{Phase 3}, exactly one of $DL$ or $DR$ will be empty, and the other set will continue to diminish in size. If for example, $DL$ is empty, elements from $DR$ will keep on getting replaced, possibly at the expense unsettling a few elements from the left-half. Thus \textsl{Phase 3} referred to as {\em asymmetric crossing phase} and will last from $(m_2+1)$ till $m_3$-th move where 
\bea
m_3 & = & \min\{ m \mid m \geq m_2, \text{ both } |DL(b(m_3))|=0 \text{ and } |DR(b(m_3))| = 0\} .\label{eq:m3} 
\eea
Since an element is never added to the set $DL \cup DR$, both symmetric and asymmetric crossing phases will terminate. At the end of asymmetric crossing phase, every value in $L(b(m_3))$ belongs to $ULR(b(m_3),t) \cup SL(b(m_3),t) \cup H(t)$. So does the case for $R(b(m_3))$. In the last \textsl{Phase 4}, settling will happen and this phase is known as {\em settling phase}. The behaviour of settling phase is determined by number of alternating cycles in $\sigma_{b(m_3)}$. While the notion of alternating cycle is developed in \cite{kumar2021oriented} for analysing routing in \SN, we redefine it in the context of \SNK. 
\bdefn  Let $c, t$ be two permutations on $[n]$ representing two distinct nodes in $\overrightarrow{S_{n,k}}$. A cycle of $c$ relative to $t$ is called alternating if it has size at least two, and elements appearing in order during a forward traversal in the cycle alternate between $ULR(c,t)$ and $URL(c,t)$. We denote by $\chi_t(c)$ the number of alternating cycles in $c$ with respect to $t$.
\edefn
When a settling move is initiated at $b(m_3)$ at the beginning  \textsl{Phase 4}, then clearly $\sigma_{b(m_3)}(1) \in \Psi_1$, where $\Psi_1$ is the cycle in $\sigma_{b(m_3)}$ containing $\sigma_t(1)$. Then there will be as many settling moves to follow as the number of elements in $\Psi_1$, finally ending at a node $r$ such that $H(r) = H(t)$. Then a seeding star-move happens so that $\sigma_t(1)$ gets added to a second alternating cycle $\Psi_2$ in $\sigma_{b(m_3)}(1)$. By this time, $ULL(c,t) \cup URR(c,t)$ will be empty, and therefore $\sigma_c(1)$ gets replaced with an element in $ULR(c,t)$ as part of seeding star-move permitting settling moves to continue unabated. Next, all values in $\Psi_2$ will get settled in the second round of settling. This will continue until all values get settled by the last move $m_L$. The number of rounds is clearly bounded by the finite quantity $\chi_t(b(m_3))$ \cite{kumar2021oriented}, and therefore \textsl{Phase 4} must also terminate with $b(m_L)=t$. Thus we conclude that the algorithm terminates correctly. 

\subsection{Analysis of the Algorithm}

Before we get on to combinatorial analysis of the algorithm, we shall establish certain results in Proposition~\ref{prop:chi-clique-move}, Proposition~\ref{prop:chi-star-move} and Proposition~\ref{prop:chi-seed-move} pertaining to number of alternating cycles newly generated while routing through nodes starting from $s$ to $t$. These numbers will matter while analysing settling phase of the algorithm. 

\bprop\label{prop:chi-clique-move} Suppose we make a clique-move at $c$ while routing a packet from $s$ to $t$ using Algorithm~\ref{RouteEven}. Let $r=\textsc{clique-move}(c,t)$. Then $\chi_t(r) \leq \chi_t(c)$.
\eprop
\bpf Let us write 
\bean
\sigma_\text{c} & = & 
	a_{\ell} \ b_2 \ \cdots  \ b_k \ a_1 \ \cdots \ a_{\ell -1} \ a_{\ell +1} \  \cdots \ a_{n-k+1}
\eean
for some fixed values of $\ell$ in $[n-k+1]$, $a_1 < a_2 < \cdots < a_{n-k+1}$ and $b_2, \ldots, b_k$. It is clear that $\sigma_r$ can be obtained as a result of composing with $\sigma_c$ two or more transpositions as defined below. First, we define  
\bea
\tau_0 = (1 \ \ k+m) \label{eq:replace}
\eea
for some $1 \leq m \leq n-k$. We may need to consider the cases $\ell \geq m$ and $\ell < m$ separately to define a subsequent set of transpositions, but without loss of generality, we may assume $\ell \geq m$. Thus we shall define
\bea
\tau_i = (k+m-i-1 \ \ k+m-i) \label{eq:sort}
\eea
for $i=0,1,\ldots, m-\ell-1$. In terms of these transpositions, we can write the relation between $\sigma_c$ and $\sigma_r$ as 
\bea
\sigma_r & = & \tau_{m-\ell-1} \circ \cdots \circ \tau_0 \circ \sigma_c \label{eq:next}.
\eea
Composition with $\tau_0$ replaces $\sigma_c(1)$ with a new value, and subsequently with $\tau_i, i>0$ sort the values $(\tau_0 \circ \sigma_c)(i), i > k$ finally to yield $\sigma_r$. For any permutation $\sigma$ and a transposition $(a \ b)$, $(a \ b) \circ \sigma$ will merge two cycles in $\sigma$ if $a, b$ fall in two distinct cycles and the merged cycle will contain both $a$ and $b$. If $a, b$ fall in the same cycle then the cycle will be split into two with one containing $a$ and the other containing $b$. By \eqref{eq:replace} and \eqref{eq:sort}, every $\tau_i, 0 \leq i \leq m-\ell-1$ is of the form $(a \ b)$ such that $a, b \in \{1,k+1,k+2,\ldots ,n\}$. Therefore, whenever one of the compositions in \eqref{eq:next} results in splitting of an existing cycle of the permutation, resultant cycles will have either  $1$ or a value greater than $k$. Thus neither of them can be an alternating cycle. In similar lines, if the composition leads to merging of two cycles, the resultant merged cycle can not be an alternating cycle. Hence $\chi_t(r) \leq \chi_t(c)$.   \epf

\bprop \label{prop:chi-star-move} Let ${\cal H'} = \{ (\sigma_c, \ \sigma_r) \mid (\sigma_c, \ \sigma_r) \in {\cal H}_{\text{star}}, (c,r) = (b(m-1),b(m)) \text{ for some } m \leq m_3 \}$ be a subset of ${\cal H}_{\text{star}}$. Then
\bean
\sum_{(\sigma_c, \ \sigma_r) \in  {\cal H'}} (\chi_t(r) - \chi_t(c)) & \leq & 1 .
\eean 
\eprop 
\bpf Without loss of generality, assume that $\sigma_c$ is an even permutation. It is clear that 
\bean
\sigma_r & = & (1 \ i) \circ \sigma_c .
\eean
for some $\sigma_c(i) \in I(t)$. If the sub-routine involves a settling move, then the cycle in $\sigma_c$ including $\sigma_c(i)$ breaks into two yielding a singleton cycle $(\sigma_c(i))$, and another cycle including $1$. Neither of them is an alternating cycle. If the sub-routing involves a crossing move from an even permutation $\sigma_c$, then $\sigma_c(i)$ can belong exclusively to one of $ULL(c,t)$, $DEL(c,t)$, $ULR(c,t)$ or $SL(c,t)$. 

If $\sigma_c(i) \in ULL(c,t) \cup  DEL(c,t)$, then $\sigma_c(i)$ can either belong to the cycle $\Psi_{\sigma_c(1)}$ or not. If it does not belong to the cycle, then the move results in a merger of two cycles. If it belongs to, then it implies that all values in $ULL(c,t) \cup  DEL(c,t)$ belong to a single cycle. In such a case, the move results in splitting of a cycle. One of the resultant cycle will include $1$, and therefore can not be an alternating cycle. The next one can turn out to be an alternating cycle only if every element in the cycle is either from $ULR(c,t)$ or from $URL(c,t)$. This further implies that $ULL(r,t) \cup DEL(r,t) = \phi$. One may jump to a conclusion that similar situation can arise for the symmetric case of $\sigma_c$ being odd, and that can generate another alternating cycle. However, we shall argue that it is not the case. Without loss of generality, assume that we hit the node $c$ first, later hit $c'$ to invoke $\textsc{star-move}(c',t)$ to move to $r'$ such that $\sigma_{c'}$ is an odd permutation and 
\bean
\sigma_{r'} & = & (1 \ i') \circ \sigma_{c'} \\
\sigma_{c'}(i) & \in & URR(c',t) \cup  DER(c',t) .
\eean 
Quite similar to the case of $r$, it follows that $|URR(c',t) \cup  DER(c',t)|=1$ and $URR(r',t) \cup DER(r',t) = \phi$. If $\sigma_{c'}(i') \in DER(r',t)$, then no more alternating cycles are generated by Proposition~\ref{prop:chi-clique-move}. If $\sigma_{c'}(i') \in URR(r',t)$, then two things can happen: first, an additional alternating cycle can be created, and second, there will be a crossing move to the left half as part of the immediate invocation of $\textsc{star-move}(r',t)$. Let $r''=\textsc{star-move}(r',t)$. Then $\sigma_{r''} = (1 \ i'') \circ \sigma_{r'}$ where $\sigma_{r'}(i'') \in ULR(r',t)$ and $\sigma_{r'}(i'')$ is part of an existing alternating cycle. As we will see soon, in such a case we shall reduce the number of alternating cycles by $1$, causing a net increase  by $1$ in the number of alternating cycles.

If $\sigma_c(i) \in ULR(c,t)$, then $\sigma_c(i)$ is chosen from an alternating cycle that does not contain $\sigma_c(1)$ if at all it exists. Then it leads to a merger of two cycles and thus reduces the number of alternating cycle by $1$. If such an alternating cycle does not exist, then $\sigma_c(i)$ can belong a non-alternating cycle. In that case, there is zero net increase in the number of alternating cycles. If $\sigma_c(i)$ is part of the cycle containing $\sigma_c(1)$, then it leads to splitting of the cycle. One of the resultant cycle will contain $1$, whereas the other can be an alternating cycle. This in turn implies that there are no alternating cycles in $\sigma_c$. Thus the net increase in alternating cycles is just $1$ in the history so far. 

Next, if $\sigma_c(i) \in SL(c,t)$, then it results in merger of a cycle with a singleton $(\sigma_c(i))$, and therefore it will not generate a new alternating cycle. This completes the proof.
\epf

\bprop\label{prop:chi-seed-move} Suppose we invoke seeding clique-move or seeding star-move at $c$ while routing a packet from $s$ to $t$ using Algorithm~\ref{RouteEven}. Let $r=\textsc{seed-clique}(c,t)$ or $r=\textsc{seed-star}(c,t)$ as the case maybe. Then $\chi_t(r) \leq \chi_t(c)$.
\eprop
\bpf The sub-routine \textsc{seed-star} or \textsc{seed-clique} is invoked only when $\sigma_c(1) = \sigma_t(1)$. Therefore, a singleton cycle $(\sigma_t(1))$ gets merged to another cycle containing $1$ when either of the two sub-routines are invoked. Therefore $\chi_t(r) \leq \chi_t(c)$
\epf 

The intention behind analysis is to compute \textsl{diam}($\overrightarrow{S_{n,k}}$) for the proposed orientation. We shall pick two arbitrary nodes $s$, $t$ and estimate an upper bound on the oriented distance between $s$ and $t$ assuming the proposed routing algorithm in Algorithm~\ref{RouteEven}. The distance is estimated by counting the number of moves that take place. We shall assume that the worst possible number of invocations indeed happen for every sub-routine. Hence the upper bound thus obtained will turn out to be an upper bound on \textsl{diam}($\overrightarrow{S_{n,k}}$) as well. In turn, \textsl{diam}($\overrightarrow{S_{n,k}}$) will be an upper bound on $\overrightarrow{\textsl{diam}}(S_{n,k})$. The final result is stated in Thm.~\ref{thm:d}.
  
\bdefn Suppose we arrive at $b(m)$ after $m$-th move while routing from $s$ to $t$ using Algorithm~\ref{RouteEven}. Then 
\bean
\alpha(m) & = &  \text{number of clique-moves}\\
\beta(m) & = & \text{number of star-moves} \\
\gamma_1(m) & = & \text{number of seeding clique-moves} \\
\gamma_2(m) & = & \text{number of seeding star-moves},
\eean
over the entire history up to the completion of $m$-th move. 
\edefn 
\bthm \label{thm:d} Let $k \geq 3, (n-k) \geq 2,$ and $k < n$. Then the oriented diameter $\overrightarrow{\textsl{diam}}(S_{n,k})$ of $\SNK$ satisfies
\bea \label{eq:dia}
\overrightarrow{\textsl{diam}}(S_{n,k}) & \leq & \left\lfloor \frac{n+k}{2} \right\rfloor + 2k + 6 - \delta(n,k) 
\eea
where $\delta(n,k)$ is non-negative function defined as
\bea \label{eq:delta}
\delta(n,k) & = & \left\{ \begin{array}{lcc} 2k-n, & & k > n/2 \\
									0, & & n/3 < k \leq n/2 \\
								   \left\lfloor \frac{n-3k}{2} \right\rfloor, & & k  \leq n/3	. 
							   \end{array}  \right. .
\eea
\ethm
\bpf Let $s$, $t$ be two arbitrary permutations. We are interested in counting $\beta(m_L), \alpha(m_L)$, $\gamma_1(m_L)$ and $\gamma_2(m_L)$ taking into the account the worst possible scenarios. 

Let us first count $\alpha(m_L)$. A clique-move happens at node $c$ if $\sigma_c(1) \in DE(s,t)$. It follows from Proposition~\ref{prop:directedge} and Lemma~\ref{lem:3cycle} that
\ben
\item[(i)] it requires one immediate clique-move routing over an edge of $Q(c)$ so that $|DI(c,t)|$ reduces by 1 if $DI(c,t) > \lfloor\frac{n-k}{2}\rfloor$;
\item[(ii)] it requires at most two subsequent clique-moves routing over two edges of $Q(c)$ so that $|DI(c,t)|$ reduces by 1 if $DI(c,t) \leq \lfloor\frac{n-k}{2}\rfloor$, except possibly in a single case when $(n-k)$ is odd that requires $3$ moves. 
\een
Let us keep the exceptional case in (ii) aside for a moment. Then for every $x \in DE(s,t)$, there can be either $1$ or $2$ clique-move(s) as decided by which of (i) or (ii) becomes valid. (You may refer to the example in Sec.~\ref{sec:ofc}.) When the first seeding clique-move happens, the output node $r$ can turn out to be such that $\sigma_r(1) \in E(t)$. This can further lead to an additional clique-move. The second seeding clique-move, if at all it happens, can not introduce any additional clique-move because the output node $r$ is such that $\sigma_r(1) \in I(t)$. This will become clearer when we analyze $\gamma_1(m_L)$. Next, let us examine when the exceptional case of $3$ moves mentioned in (ii) can happen. When $(n-k)$ is odd, by Lemma~\ref{lem:3cycle}, we may end up routing via the lone $4$-cycle of an oriented fundamental clique.(You may again refer to the example with $(n-k)=5$ in Sec.~\ref{sec:ofc}) Since path with the shortest length is preferred, the case of routing via a path of length $3$ happens only when there is a single displaced internal value left out. Thus in the whole history of moves, one additional clique-move can happen if $(n-k)$ is odd. Let $m_{\text{zd}} \leq m_3$ be the least integer such that $|DI(b(m_{\text{zd}}),t)| = 0$. Then
\bea \label{eq:pmzd}
\begin{aligned}
\alpha(m_{zd}) \leq {} &  \max\left\{ DE(s,t) - \left\lfloor\frac{n-k}{2}\right\rfloor , 0 \right\} + 2\min\left\{ DE(s,t), \left\lfloor\frac{n-k}{2}\right\rfloor  \right\} \\&  \hspace{2in} \ + \ 2  
\end{aligned} 
\eea
The node $b(m_{\text{zd}})$ belongs to the star-subgraph of destination $t$, and there shall be no more clique-moves thereafter. Therefore 
\bea 
\alpha(m_L) & = & \alpha(m_{\text{zd}}). \label{eq:p}
\eea 

Next we move on to count $\beta(m_L)$. We shall first count star-moves that involve a crossing move. We shall argue that for every $x \in ULL(s,t) \cup DEL(s,t)$, there will be two uniquely associated crossing moves. Let the pair of moves be   $(m_{\text{c}}(x), m'_{\text{c}}(x))$ with $m_{\text{c}}(x) < m'_{\text{c}}(x)$. The  move at $m_{c,1}(x)$ is characterised by the fact that $\sigma_{b(m_{\text{c}}(x))}(1)$ becomes equal to $x$. The second one depends on whether $x \in DEL(s,t)$ or $x \in ULL(s,t)$. If $x \in DEL(s,t)$, then $m'_{\text{c}}(x)$-th star-move will follow after at most two clique-moves such that $\sigma_{b(m'_{\text{c}}(x)-1)}(1) \in DI(s,t)$. If $x \in ULL(s,t)$, then $m'_{\text{c}}(x)=m_{\text{c}}(x)+1$. This move can be a settling move, but we shall assume it as a crossing move to account for the worst case. A similar argument holds true for every element in $URR(s,t) \cup DER(s,t)$. However, if we are in symmetric crossing phase, two observations are in place: first, the $m_{\text{c}}(x)$-th move turns out to be the $m'_{\text{c}}(y)$-th move associated to an element in $y \in URR(s,t) \cup DER(s,t)$; second, the $m'_{\text{c}}(x)$-th move turns out to be the $m_{\text{c}}(y')$-th move associated to an element $y' \in URR(s,t) \cup DER(s,t)$. Thus we can so far account for a total of \bean
2 \min\{|ULL(s,t) \cup DEL(s,t)|, |URR(s,t) \cup DER(s,t)|\}
\eean
star-moves. Again for the remaining 
\bean
\begin{aligned}
 & \max\{|ULL(s,t) \cup DEL(s,t)|, |URR(s,t) \cup DER(s,t)|\} \ - \\
 & \hspace{1in} \min\{|ULL(s,t) \cup DEL(s,t)|, |URR(s,t) \cup DER(s,t)|\}
\end{aligned}
\eean
elements, there will be $2$ associated star-moves in asymmetric crossing phase adding up to a total of  
\bean 
2 \max\{|ULL(s,t) \cup DEL(s,t)|, |URR(s,t) \cup DER(s,t)|\}
\eean 
crossing moves.

Next, we shall count the remaining star-moves in all phases in order. Clearly, there shall be no crossing move in the transient phase. Every single settling move, if any, will settle one unsettled value in $I(t)$. We have already counted crossing moves happening in both symmetric and asymmetric crossing phases. A displaced internal value can get settled in the \textsl{Phase 2}, but we have assumed that every displaced internal value gets subjected to a crossing move in order to account for the worst case. In \textsl{Phase 3}, values can get settled consuming a single move. Quite distinctly in \textsl{Phase 3}, an already settled value can turn unsettled at most 
\bean
\begin{aligned}
& \max\{|ULL(s,t) \cup DEL(s,t)|, |URR(s,t) \cup DER(s,t)|\} \ - \\ 
& \hspace{1in} \min\{|ULL(s,t) \cup DEL(s,t)|, |URR(s,t) \cup DER(s,t)|\}
\end{aligned}
\eean
times as part of crossing moves. This may presumably prompt one to count for multiple settling moves for the same element. We shall argue that this is not required. Suppose we are at node $c$, and the next move turns a settled value $x \in SL(c,t)$ into an unsettled one. This means that
\bean 
|ULL(s,t) \cup DEL(s,t)| & > & |URR(s,t) \cup DER(s,t)|.
\eean
We shall imagine an alternate source $s'$ such that 
\bean 
|ULL(s,t) \cup DEL(s,t)| - |URR(s,t) \cup DER(s,t)|
\eean 
additional values are present in $ULL(s',t)$ so that both the counts are balanced in $s'$. And under this imaginary situation, there shall never be a case of a settled value turning unsettled. The point however is that the estimate made on the count of star-moves remain the same in both the real case of $s$ and the imaginary case of $s'$. Thus we shall assume that no settled value gets unsettled. Finally, let us consider the settling phase. In this phase, one may observe that there is a numerical symmetry as given by
\bea
||SL(b(m_3),t)| - |SR(b(m_3),t)| | & \leq & 1 \label{eq:ns1} \\
||ULR(b(m_3),t)| - |URL(b(m_3),t)| | & \leq & 1 \label{eq:ns2},
\eea
where the discrepancy factor of $1$ in size may possibly arise due to $\sigma_t(1)$. Suppose
\bea \label{eq:z}
\begin{aligned}
	 ULL(b(m_3-1),t) \cup DEL(b(m_3-1),t) \cup URR(b(m_3-1),t) & \\ \hspace{1in} \cup \ DER(b(m_3-1),t) & =  \{z \} 
\end{aligned}
\eea  
and without loss of generality let us say $z \in ULL(b(m_3-1),t) \cup DEL(b(m_3-1),t)$. Clearly, the $m_3$-th move is a star-move that replaces $z$. If $z \in ULL(b(m_3-1),t)$, then in $(m_3+1)$-st move, $z$ will replace an element in $URL(b(m_3),t)$, and this will be the last crossing move. All the subsequent moves will be settling or seeding star-moves. On the other hand, if $z \in DEL(b(m_3-1),t)$, then within one or two subsequent clique-moves we shall arrive at a node $c$ within destination star-subgraph. If $\sigma_{c}(1) = \sigma_t(1)$, then what follows is a seeding star-move (unless destination is already reached), and subsequent moves will be either settling or seeding moves. If $\sigma_{c}(1) \neq \sigma_t(1)$, then what follows can be a crossing move depending on the sign of $c$. This will be the last crossing move, and all subsequent moves will be settling or seeding star-moves. Every element $x \in ULR(b(m_3),t)$ can be associated with two star-moves before it gets settled. After the first, the resultant node $r$ will have $\sigma_r(1) = x$, and after the second, the resultant node $r'$ will have $\sigma_{r'}^{-1}(x) = \sigma_{t}^{-1}(x)$. However, these two moves will also witness settling of an element $y \in URL(b(m_3),t)$. Remember that there is a numerical symmetry between $|ULR(b(m_3),t)|$ and $|URL(b(m_3),t)|$. Thus we shall have one settling move per every element in $ULR(b(m_3),t)$. In addition, a settled value other than $\sigma_t(1)$ will not turn unsettled after the $m_3$-th move, as this can only happen as part of a crossing move raising a contradiction. Thus we conclude that there will be at most $k$ settling moves, counting one each for every element that gets settled. Thus we have
\bea
\beta(m_L) & \leq & 2 \max\{|ULL(s,t) \cup DEL(s,t)|, |URR(s,t) \cup DER(s,t)|\} + k.  \label{eq:q}
\eea

Let us proceed to count $\gamma_1(m_L)$ and $\gamma_2(m_L)$. A seeding clique-move can happen if $\sigma_c(1)=\sigma_t(1)$ and $DEA(c,t) > 0$, i.e., $IAT(c,t)=IA(t)\cap T(c)$ is non-empty. Only if a clique-move subroutine enters {\em Case 2.2}, the possibility of a second seeding clique-move arises. The case can occur only after $m_{\text{sd}}$-th move where $m_{\text{sd}} < m_L$ is the least integer such that
\bean 
|IAT(b(m_{\text{sd}}),t)| & = & 1.
\eean
If the routing algorithm hits {\em Case 2.2} of Algorithm~\ref{CliqueMove}, then a seeding clique-move will happen within two subsequent moves if $(n-k)$ is even or at most three subsequent moves if $(n-k)$ is odd (by virtue of the lone $4$-cycle) to reach a node $r$. Clearly $|DEA(r,t)| = 0$, and hence there shall not be any more seeding clique-move. Thus
 \bea
\gamma_1(m_L) & \leq & 2 \label{eq:r1}.
\eea 
A seeding star-move happens when we are at a node $c$ with $\sigma_c(1) = \sigma_t(1)$ and $DEA(c,t) = 0$. If $r$ is the output node after the first seeding star-move, then a numerical symmetry as given below
\bea
||ULL(r,t)\cup SL(r,t)| - |URR(r,t)\cup SR(r,t)| | & = & 1 \label{eq:ns3} \\
||ULR(r,t)| - |URL(r,t)| | & = & 1 \label{eq:ns4},
\eea
must hold true. Let $\sigma_r(1)=x$. If the move happens in one among the first three phases, then it is possible that $x \in ULL(c,t)$, $x \in SL(c,t)$ or $x \in ULR(c,t)$ picked in this order of preference. Whichever may be the case, we shall argue that there can not be a second seeding star-move in first three phases. If $x \in ULR(c,t)$, then it must be that $ULL(c,t) \cup SL(c,t) = \phi$. By \eqref{eq:ns3} and \eqref{eq:ns4}, this means that we have already got on to the fourth phase as part of the first seeding star-move. If $x \in ULL(c,t) \cup SL(c,t)$, then the second seeding star-move is feasible only as part of settling of $\sigma_c^{-1}(x)$. The settling move for $\sigma_c^{-1}(x)$ can however happen only in the fourth phase. Thus we conclude that there is at most one seeding star-move in the first three phases. By definition, $b(m_3)$ is the last node visited before settling phase starts, and it is clear from \eqref{eq:z} that $\sigma_{b(m_3)}(1)=z$. Let $\chi_0 = \chi_t(b(m_3))$ and ${\cal C} = \{c' \mid \sigma_{c'}(1)=\sigma_t(1), \text{ node }c' \text{ is visited after } b(m_3) \}$. If $z \in DEL(b(m_3-1),t)$ and $\sigma_t(1) \in DI(b(m_3),t)$, then it is quite possible that we hit a node $c$ in the destination star-subgraph with $\sigma_c(1)=\sigma_t(1)$ before settlings start. In that case $|{\cal C}|=\chi_0+1$. Furthermore, there must not have been any seeding star-move in the first three phases. On the other hand if $z \in ULL(b(m_3-1),t)$, first round of settlings will happen before we hit a node $c$ such that $\sigma_c(1)=\sigma_t(1)$. In that case, $|{\cal C}|=\chi_0$. Thus we conclude
\bean
\gamma_2(m_L) & \leq & 1 + \chi_0 .
\eean
Then it follows from Proposition~\ref{prop:chi-clique-move},  Proposition~\ref{prop:chi-star-move} and Proposition~\ref{prop:chi-seed-move} that
\bean
\gamma_2(m_L) & \leq & 1 + \chi_t(s) + \sum_{(\sigma_c,\sigma_t) \in {\cal H}_{\text{star}}} \chi_t(r) - \chi_t(c) \nonumber \\ & &+  \sum_{(\sigma_c,\sigma_t) \in {\cal H}_{\text{clique}}} \chi_t(r) - \chi_t(c) + \sum_{(\sigma_c,\sigma_t) \in {\cal H}_{\text{seed}}} \chi_t(r) - \chi_t(c) \\
& = & 2 + \chi_t(s) .
\eean
Assuming the worst case scenario, we shall assume that $ULR(s,t)=URL(s,t)=\phi$, and therefore $\chi_t(s)=0$. It follows that 
\bea \gamma_2(m_L) & \leq & 2 \label{eq:r2}
\eea 
Combining \eqref{eq:p}, \eqref{eq:q}, \eqref{eq:r1} and \eqref{eq:r2}, we obtain:
\bea
\overrightarrow{d}(s,t) & \leq & \alpha(m_L) + \beta(m_L) + \gamma_1(m_L) + \gamma_2(m_L) \nonumber \\
& = & \max\left\{ DE(s,t) - \left\lfloor\frac{n-k}{2}\right\rfloor , 0 \right\} + 2 \min\left\{ DE(s,t), \left\lfloor\frac{n-k}{2}\right\rfloor  \right\} \nonumber \\
& & \hspace{.2in} + \ 2 \max\{|ULL(s,t) \cup DEL(s,t)|, |URR(s,t) \cup DER(s,t)|\}  \nonumber \\  & & \hspace{.2in} + \ k  + 5 \label{eq:maincount}
\eea
We proceed to bound the sizes of relevant sets in \eqref{eq:maincount} considering the worst case. 
The term $\max\left\{ DE(s,t) - \left\lfloor\frac{n-k}{2}\right\rfloor , 0 \right\}$ exceeds zero only when $k > n/3$. Moreover, $|DE(s,t)| = \min\{k,n-k\}$ in the worst case. Substituting these worst-case estimates, we can compute 
\bean
\max\left\{ DE(s,t) - \left\lfloor\frac{n-k}{2}\right\rfloor , 0 \right\} + 2 \min\left\{ DE(s,t), \left\lfloor\frac{n-k}{2}\right\rfloor  \right\} & = & \left\lfloor \frac{n+k}{2} \right\rfloor - \delta(n,k) 
\eean
where $\delta(n,k)$ is as defined in \eqref{eq:delta}. It is easy to observe that
\bean
2 \max\{|ULL(s,t) \cup DEL(s,t)|, |URR(s,t) \cup DER(s,t)|\} & \leq & k .
\eean
Putting these together yields
\bean
\overrightarrow{d}(s,t) & \leq & \left\lfloor \frac{n+k}{2} \right\rfloor + 2k + 6 - \delta(n,k).
\eean


\epf

\section{Comparison and Future Directions}

We provide a comparison of various bounds on oriented diameter of an $(n,k)$-star graph in Table~\ref{Table1}.  
\begin{table}[ht]
	\centering
	\begin{tabular}{ |c|c|c|}
		\hline
		\multirow{2}{3cm}{\centering{Cheng and Lipman~\cite{cheng2002unidirectional}}}&$10k-5$&\text{when }$1 \leq k \leq \lfloor \frac{n}{2}\rfloor$\\
		\hhline{~--}&$5k+5 \lfloor \frac{(n-1)}{2}\rfloor$&\text{when }$\lfloor \frac{n}{2}\rfloor+1 \leq k \leq (n-1)$\\
		\hline
		\hline
		
		\multirow{2}{3cm}{\centering{Cheng and Kurk~\cite{cheng2006routing}}}&$6(k-3)+13$&\text{when }$(n-k)$\text{ is odd}\\
		\hhline{~--}&$7(k-3)+18$&\text{when }$(n-k)$\text{ is even}\\
		\hline
		\hline
		\multirow{3}{3cm}{\centering{Our Result}}&{\centering{$\left\lfloor \frac{n+k}{2} \right\rfloor + 2k + 6 - \delta(n,k)$}}& $\delta(n,k) = 2k-n$\text{ when } $k>\frac{n}{2}$\\
		\hhline{~~-}&&$\delta(n,k) = 0$\text{ when } $\frac{n}{3} < k\leq \frac{n}{2}$\\
		\hhline{~~-}&&$\delta(n,k) = \lfloor \frac{n-3k}{2}\rfloor$\text{ when } $k\leq \frac{n}{3}$\\
		
	\hline
	\end{tabular}
	\vspace*{2mm}
	\caption{Comparision table of results for the corresponding upper bounds on the diameter of $\protect\USNK$}\label{Table1}.
\end{table}
The upper bound derived in the present work dominates all existing bounds. This becomes evident if we write bounds in Table~\ref{Table1} in terms of $k$ alone by substituting the upper bound on $n$ in every range considered. The bound in \cite{cheng2002unidirectional} becomes $\overrightarrow{d}(s,t) \leq 10k-5$ if $1 \leq k \leq \lfloor (n/2) \rfloor$ and $\overrightarrow{d}(s,t) \leq 10k-5 - 5(n+1 \mod 2)$ if $\lfloor (n/2) \rfloor < k \leq n-1$. The bound is \cite{cheng2006routing} is already in terms of $k$ alone. Our bound becomes 
\bea \label{eq:kdependence}
\overrightarrow{d}(s,t) & \leq & \left\{ \begin{array}{lc} \lfloor 3.5k \rfloor +6, & n < 2k \\
	4k + 6, & n \geq 2k \end{array}. \right.
\eea
A comparison of the dependence on $k$ in \eqref{eq:kdependence} with that of bounds in \cite{cheng2002unidirectional},\cite{cheng2006routing} helps to evaluate the dominance of our bound.

Both the proposed orientation and the proposed distributed routing algorithm match exactly with that in \cite{kumar2021oriented} within star-subgraphs of \SNK. Thus it becomes meaningful to compare our bound for \SNK~with that for $S_k$ provided in \cite{kumar2021oriented}. In \cite{kumar2021oriented}, $\textsl{diam}(\orient{S_{k}})$ is shown to be bounded above by $2k+4$. In the worst case of $k\leq n/3$, the two bounds are separated by a gap of $2k+2$, while in the best case of $k=n-1$ they match except for an additive constant. The gap varies as a function of $k/n$, and it is a natural question to investigate whether such a gap is unavoidable for the particular orientation. A second question pertains to how well the upper bound on $\overrightarrow{\textsl{diam}}(S_{n,k})$ compares with the undirected diameter $\textsl{diam}(S_{n,k})$, a natural lower bound for the oriented diameter. When $k \leq n/3$, the gap is pegged at $2k+7$, and as $k$ increases from $\lceil n/3\rceil$, the gap reduces to hit $\lfloor 0.5k\rfloor + 6$ at $k=n-1$. The crux of answering these questions lies in coming up with good lower bounds on directed diameter of $\orient{S_{n,k}}$.

 \bibliographystyle{splncs04}
 \bibliography{nkstar}

\end{document}